\documentclass[12pt]{article}
\usepackage[dvips]{graphics}
\RequirePackage{ifthen}[1994/05/27]
\newboolean{drafton}    \setboolean{drafton}{false}
\ifthenelse{\boolean{drafton}}%
{    \newcommand{\equn}[1]{
        \mbox{}~\hspace*{\stretch{1}}~\begin{picture}(0,0)
           \setlength{\fboxsep}{.7mm}
           \put(3,-12){\makebox[0mm][l]{\fbox{\small #1}}}
        \end{picture}
        \begin{equation}\label{#1}}
    \newcommand{\eqan}[1]{
        \setlength{\fboxsep}{.7mm}
        \mbox{}~\hspace*{\stretch{1}}~\begin{picture}(0,0)
        \put(3,-12){\makebox[0mm][l]{\fbox{\small #1}}}
        \end{picture}
        \begin{eqnarray}\label{#1}}}
{   \newcommand{\equn}[1]{\begin{equation}\label{#1}}
    \newcommand{\eqan}[1]{\begin{eqnarray}\label{#1}}}
\newcommand{\bosy}[1]{\mbox{\boldmath $#1$}}
\newcommand{\eqa}{\begin{eqnarray}}
\newcommand{\equ}{\begin{equation}}
\newcommand{\nuqe}{\end{equation}}
\newcommand{\uqe}{\end{equation}}
\newcommand{\naqe}{\end{eqnarray}}
\newcommand{\aqe}{\end{eqnarray}}
\newcommand{\nonu}{\nonumber}

\newcommand{\goto}{\rightarrow}
\newcommand{\half}{\frac{1}{2}}

\newcommand{\n}{{\bf \nabla}}
\newcommand{\e}{{\rm e}}
\newcommand{\Si}{\sigma}
\newcommand{\ka}{\kappa}
\newcommand{\q}{{\bf q}}
\newcommand{\y}{{\bosy \rho}}
\newcommand{\dH}{\delta h}
\newcommand{\rw}{{\rm s}}

\begin{document}

\title{The Influence of Substrate Structure on Membrane Adhesion} 
\author{Peter S.\ Swain$^{(1,2)}$ and David Andelman$^{(2)}$ 
\\
\\
$^{(1)}$Max-Planck Institut f\"{u}r Kolloid- und
Grenzfl\"{a}chenforschung,\\ 14424 Potsdam,
Germany
\\
\\
$^{(2)}$School of Physics and Astronomy,\\
Raymond and Beverly Sackler Faculty of Exact Sciences,\\ 
Tel Aviv University, Ramat Aviv 69978, Tel Aviv, Israel}
\date{}
\maketitle

\begin{abstract}
We consider a membrane both weakly and strongly adhering to a
geometrically structured substrate. The interaction potential is assumed
to be local, via the Deryagin approximation, and harmonic. Consequently,
we can analytically describe a variety of different geometries: as well
as randomly rough self-affine surfaces, smooth substrates interrupted by
an isolated cylindrical pit, a single elongated trench or a periodic
array of trenches are investigated. We present more general expressions
for the adhesion energy and membrane configuration in Fourier space and
find that, compared to planar surfaces, the adhesion energy decreases. 
We also highlight the possibility of overshoots occurring in the membrane
profile and look at its degree of penetration into surface indentations.
\end{abstract}



\pagebreak
\section{Introduction}
\setcounter{equation}{0}

The statistical mechanics of membranes is an important branch of
soft-condensed matter physics, not least because of its application to
biological systems. Examples of membranes that can be studied
experimentally include liposomes or vesicles, microemulsions, lamellar
liquid crystals, as well as biological cells, such as red blood cells
\cite{sackmann}. Of great importance is a detailed understanding of the
adhesion between two membranes or between a membrane and a substrate. 
This occurs ubiquitously in nature, with vesicle adhesion playing a
dominant role in endo- and exocytosis \cite{alberts}, that is the
communication of a cell with its immediate environment. Efficient drug
delivery is dependent on the adhesion between a liposome and the plasma
membrane of the target site \cite{lasic} while adhesion phenomena are
also indispensable to biotechnology with, for example, biosensors being
based on the binding of membranes to substrates.

In this paper we choose to concentrate on the latter, that is the
adhesion between a membrane and a solid substrate, in order to provide
theoretical support for recent experiments aimed at creating new
biotechnology. All of this research has involved the study of adhesion on
materials which are not flat and chemically homogeneous but that have a
deliberate chemical or geometrical patterning. As far as we are aware,
there has been no theoretical work describing membrane adhesion on such
structured surfaces. Previous studies have looked at vesicle adhesion on
simple, planar, chemically homogeneous substrates \cite{seiflip,seif} and
we hope now to redress this imbalance. 

First of all, we distinguish between weakly and strongly adhering
vesicles or membranes. Weakly adhering giant vesicles have been studied
using reflection interference contrast microscopy \cite{rad} and have
been found to lie at distances between $300 \, {\mbox \AA}$  and $400 \, {\mbox \AA}$ from a planar substrate. Here, Van der Waals attractions to the
substrate are mainly counterbalanced by a repulsive entropic Helfrich
force. Strongly adhering or supported membranes \cite{sack}, however,
sandwich a water or polymer film between them and the substrate and
typically have a much smaller separation lying at distances between $10
\, {\mbox \AA}$ and $40 \, {\mbox \AA}$. Hydration forces are now the
dominant repulsive interaction. 

Supported membranes provide perhaps the most potential for
biotechnological applications. They can be formed by the spreading of a
bilayer over a substrate, vesicle fusion taking place at a substrate or
by lipid monolayer transfer using a Langmuir-Blodgett technique
\cite{sack}. These membranes are useful as they enable
biofunctionalization of inorganic solids and provide a means to
immobilize proteins, e.g.\ lipid coupled antigens and antibodies, with a
well defined orientation and in non-denaturing conditions \cite{sala}.
This then offers a suitable environment to investigate protein-membrane
coupling and protein-protein recognition processes. They also can be used
to design phantom cells which allow study of the interplay between
specific (lock and key) and universal forces during cell adhesion
\cite{sack}. 

However, we wish to highlight the ability of supported membranes to act
as biosensors. They can provide ultra-thin, highly electrically resistant
layers on top of a conducting substrate. If protein receptors are
incorporated into these layers then electrical or optical detection of
ligand binding to the receptors creates a biosensor.  Alternatively,
protein ion channels can be embedded in the membrane in which case their
effect on its permeability endows the membrane with sensor-like
properties.

Proteins included in the membrane often have a dimension greater than its
thickness (which is approximately $40 {\mbox \AA}$) and so can lift the
membrane off the substrate. This is undesirable and can be prevented by
creating water pockets on the surface of the substrate which then act as
protein ``docking'' sites. Therefore, one is naturally led to consider
membrane adhesion on geometrically structured substrates which have been
indented in some way. This structure can be made by a variety of methods,
for example, chemical etching of silicon wafers can lead, due to their
particular crystal structure, to long ``V''-shaped channels
\cite{gerdes}. Alternatively, one can use micelles made from diblock
copolymers to coat any semiconductor wafer. If this is ion
 sputtered, then loading of the micelles before hand with an inorganic
compound can provide large etching contrasts. Using this technique one can
pattern a planar substrate with pits or islands on a nanometer scale 
\cite{moller}. 

Motivated by such experiments, we discuss here a straightforward
theoretical approach which allows the adhesion energy of a bound membrane
to be calculated for a variety of different geometrical configurations of
the substrate. The techniques used have been strongly influenced by
recent work in wetting phenomena. There is now quite a substantial
literature covering the effect of geometrical and chemical substrate
structure on wetting films. A general form of Young's equation has been
derived \cite{marmur,swalip}, the influence of disorder \cite{dis} and
corrugation \cite{sp} of the substrate on wetting transitions determined
and the effect of chemical patterning \cite{peter} on the shapes of
adhering droplets studied. However, we use as the inspiration for our
approach, a systematic description of the configuration of wetting films
on non-planar surfaces \cite{david,dr,david2} in which the full non-local
form of the Van der Waals interaction can be accounted for. 

There are also many interesting applications involving the adhesion of
membranes on chemically structured substrates. However, we will not
discuss this further here but return to it again in an accompanying
publication \cite{II}. 

To start, we describe the various molecular interactions included in our
model and differentiate between strong and weak adhesion. In Sec.\
\ref{aplanar}, we consider simple, planar surfaces, which serve as basis
for nearly all the following work. The most profitable analytic approach,
the Deryagin approximation, is detailed in Sec.\ \ref{dery}. We consider
(separately) a corrugated substrate (Sec.\ \ref{sinusoidally}) and one
broken by a single (long) trench (Sec.\ \ref{anisol}), a pit (Sec.\
\ref{anisolpit}) and a periodic array of trenches (Sec.\ \ref{trsSec}).
We calculate the equilibrium membrane profile and the adhesion energy.
The variation of this energy with respect to the parameters
characterizing the geometric structure is emphasized. In Sec.\
\ref{selfaffine}, the adhesion energy for randomly rough or self-affine
surfaces is given for two particular cases. Finally, we conclude with a
short discussion.


\section{The Free Energy}
\label{thefree}
\setcounter{equation}{0}

\begin{figure}
\begin{center}
\scalebox{0.6}{\includegraphics{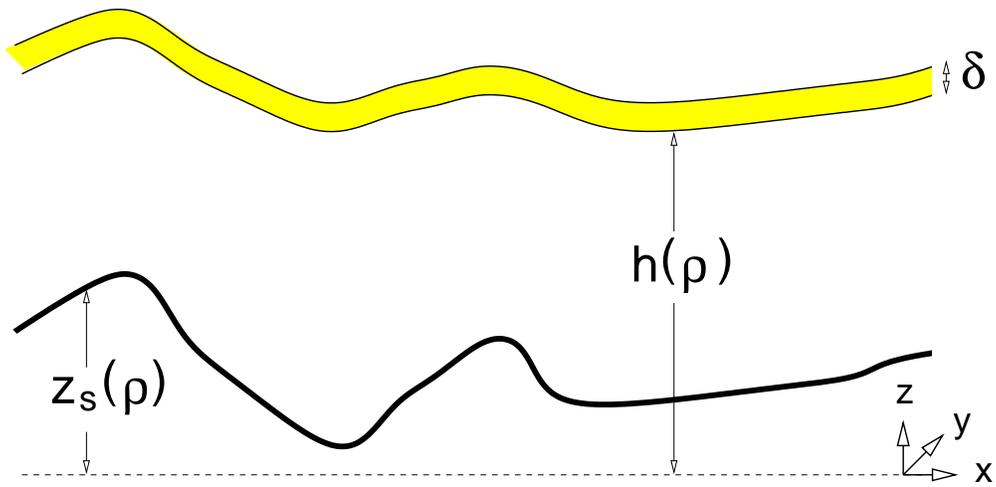}}
\end{center}
\caption{A membrane adhering to a rough surface. The reference $\y$-plane
is shown as a dashed line. The height of the lower membrane lipid leaflet
and the surface, measured from this plane, are denoted by $h(\y)$ and
$z_\rw(\y)$, respectively. The membrane thickness is $\delta$.}
\label{fig1}
\end{figure}

To begin, we consider a membrane with an elastic modulus $\ka$ and
 tension $\Si$ interacting with a rough surface. For a free,
almost flat membrane, which is infinite in extent, the bending energy can
be described by an effective free energy functional \cite{canham}
\equ 
\int ds_1 ds_2 \, \sqrt{g} \, \left[ \sigma + \half \kappa (2H)^2
\right] 
\uqe 
where $(s_1,s_2)$ are coordinates in the membrane surface, $g$ is the
determinant of the metric, $H=(c_1+c_2)/2$ the mean curvature, and $c_1$
and $c_2$ the two principal curvatures.  As the Gaussian curvature is a
total divergence, the Gauss-Bonnet theorem implies that it can be ignored
for a membrane with fixed topology and we proceed to do just this here.
Working in the Monge representation, let $\y = (x,y)$ be a two
dimensional planar vector and the heights of the surface and membrane
above some reference $\y$-plane be $z_\rw(\y)$ and $h(\y)$, respectively.
The geometry and notation used is shown in Fig.\ \ref{fig1}. 

For an experimental system, in which giant vesicles (rather than the
infinite membrane considered here) are involved, the tension usually
arises due to the conservation of the vesicle total surface area.
However, an additional contribution develops, for a weakly adhering
vesicle, as the lipids making up the membrane try and move from areas far
from the substrate into the more energetically favorable region near the
surface. However, for our case of an infinite membrane none of these
intricacies arise and we can consider the tension as an external
parameter. 

If the membrane thickness is $\delta$, the interaction between the
surface and the membrane can be accounted for by a potential
$V(h,\delta; z_\rw)$ and total membrane free energy is given as
\equn{f}
{\cal F}[h] = \int d\y \Biggl \{ \sqrt{1+(\n h)^2} \left[ \Si +
\frac{\ka}{2} \left( {\bosy \n} \cdot \frac{{\bosy \n} h}{\sqrt{1+(\n
h)^2}} \right)^2 \right] + V(h,\delta;z_\rw) \Biggr \}
\nuqe
where we have used the Monge representation.

The interaction can be broken down into several components
\cite{lip,isr}. We will always include an external pressure, a Van der
Waals attraction and a repulsive force, whose nature will be different
for strong and weak adhesion. In principle, other possible interactions,
for example electrostatic forces, can be added for more refined
calculations.
The total potential, used in (\ref{f}), is then
\equn{nochein}
V(h,\delta; z_\rw) = p \cdot (h-z_\rw) + V_{\rm vdw}(h,\delta; z_\rw) 
+ V_{\rm rep}(h; z_\rw) 
\nuqe 
A different dependence on the surface height has been emphasized as, in
general, the interaction potential is a {\it functional} of $z_\rw$. Note
that in (\ref{f}), the potential $V$ is integrated over the 
{\em projected area}, $A_0=\int d\y$, only because all non-local
effects of the solid roughness and membrane fluctuations are, in principle,
incorporated into the potential itself.

The existence of an external pressure $p$ in (\ref{nochein}) implies that
there is a difference in pressure across the membrane. Recalling that the
membrane is infinite in its lateral extent, the pressure couples linearly
to the membrane height and could arise, for example, from the existence
of macromolecules on one side of the membrane only. An additional
contribution arises due to the finite membrane thickness but as we assume
that this is fixed throughout the membrane, it is just a
constant which we neglect. 

The Van der Waals attraction that the membrane feels towards
the solid surface is given by 
\equn{pot0}
V_{\rm vdw}(h,\delta;z_\rw) = - \Bigl[ W(h;z_\rw) - W(h+\delta;z_\rw) \Bigr]
\nuqe
where $W(h;z_\rw)$ is an interfacial-like Van der Waals potential.
Equation (\ref{pot0}) arises due to the bilayer nature of the membrane
and can be understood as a supposition of two wetting layers. One of
these has height $h+\delta$ while the other is of height $h$ and has a
Hamaker constant of opposite sign. For a wetting fluid film of
thickness $h(\y)$ resting on a solid substrate, with a rough surface
configuration given by $z=z_\rw(\y)$, $W(h;z_\rw)$ satisfies \cite{david}
\equn{pot}
W(h;z_\rw) = \int_{h(\y)}^\infty dz \int d\y' \int_{-\infty}^{z_\rw(\y+\y')}
dz'\,\, w({\y}',z-z') 
\nuqe
with a kernel interaction
\equ
w({\bf r}) = \frac{A}{\pi^2} \left( \frac{1}{r} \right)^{2m+2}
\uqe
Following convention, we have denoted the Hamaker constant by $A$ and
will usually set the integer $m=2$ to model non-retarded Van der Waals
interactions. Equation (\ref{pot}) comes from the sum over all possible pair
interactions between the molecules making up two half spaces which are
each capped by either the surface $z=h(\y)$ or $z=z_\rw(\y)$.

Notice that (\ref{pot}) is a function of $h(\y)$ but a functional of
$z_\rw(\y)$. For a flat membrane on top of a smooth and planar solid
surface, which will be discussed in the following section, this
functional dependence on $z_\rw(\y )$ can be ignored and then (for $m=2$) 
\equ
W(h,z_\rw) = \frac{A}{12\pi} \cdot \frac{1}{(h-z_\rw)^2}
\uqe
This is the standard result for the Van der Waals potential between two
semi-infinite bodies with planar surfaces held a distance $h-z_\rw$ apart
\cite{isr}. 

To counterbalance this attractive force, a repulsive interaction is
included and, as before, we define two different possible scenarios; the
weak and strong adhesion regimes:

\subsection{Weakly adhering membranes}

In this case we consider a repulsive force which is entropic in origin. 
This arises due to the surface cutting off the region that can be sampled
by the membrane as it undergoes thermal fluctuations.  A renormalization
group description is the only general approach that can be used to
describe such effects but, providing fluctuations do not become critical,
a simple supposition ansatz can suffice \cite{lip}. In this, so-called
weak fluctuation regime \cite{lipfish}, an entropically based term is
added to the membrane potential. Such a term was first described by
Helfrich \cite{helf2} for a membrane (with zero tension) in a stack of
membranes and can be thought of as an ideal ``gas'' pressure. The
presence of another membrane above and below introduces a new lengthscale
describing the maximum wavelength at which any membrane within the stack
can fluctuate without being strongly suppressed by its neighbors.
Fluctuations with wavelengths below this lengthscale are essentially
unhampered and so one can think of an ideal gas of uncorrelated membrane
humps characterized by a single size. The pressure that this gas exerts
is simply the Helfrich repulsion. 

When a membrane is under tension, the extent to which it can fluctuate is
much reduced. Therefore, in some sense, the membrane ``feels'' the
presence of a surface or another membrane less and there is a
corresponding decrease in entropic repulsion. This effect was looked at
by Helfrich and Servuss \cite{helfserv}, who, within a mean-field theory,
found that the entropic repulsion, between two membranes, decayed
exponentially in $h^2$ in the tension-dominated regime. Such a result has
also been confirmed by Evans \cite{eva} via a self-consistent mean-field
method. However, by using a functional renormalization group approach
\cite{lip} to include thermal fluctuations, it has been argued that this
decay, while remaining exponential, should be so in $h$ only. This has
been verified by Monte-Carlo simulation \cite{rolandlip}. Analytical
forms of the entropic repulsion are only available for the tension and
rigidity dominated regimes but ideally one would like to have a function
which tends to these two expressions in the appropriate limits. While
this function has been found numerically \cite{rolandlip}, there does not
yet exist a complete analytical expression for it. However, a simple
argument by Seifert \cite{seif} gives a form of the entropic interaction
which takes the correct limit for $\Si \goto 0$ or $h \goto 0$ and has
the dominant exponential decay in $h$ also present in the tension regime.
We choose, for simplicity, to adopt this form which is given by

\equn{helf}
V_{\rm fluc}(h) = \frac{6 T^2}{(2\pi)^2 \kappa } \left(
\frac{\Omega }{\sinh \Omega h} \right)^2
\nuqe
where
\equ
\Omega = \left(\frac{\pi\Si}{2T} \right)^\half 
\uqe

We denote the temperature by $T$ and have set here and hereafter the
Boltzmann constant $k_B$ to unity. The prefactor of (\ref{helf}) has been
chosen to be consistent with the renormalization group description
\cite{lip} as $h \goto \infty$ but we note that the exact value has no
special importance in the model presented here.  In the limit of zero
tension ($\Si/ka \sim \Si/T \goto 0$), (\ref{helf}) tends to the well
known result ${3T^2}/(2\pi^2 \ka h^2)$. For large $\Omega h$, the
potential decays exponentially as $\exp(-2\Omega h)$.  In this limit,
which can correspond to large $\Si$, the membrane should take on more
interfacial-like behavior and, indeed, the entropic repulsion for an
interface in three dimensions is also expected to decay exponentially
with height \cite{lipfish}. 

As $\delta$ is constant we can ignore its influence on the fluctuation
interaction; it is only the lower lipid leaflet which defines the volume
between the membrane and the substrate and so determines $V_{\rm fluc}$. 

Equation (\ref{helf}) was initially assumed only to be valid for a
membrane fluctuating near a flat surface. However, a number of different
methods have shown that the Helfrich term can also be used for a rough
substrate if $h$ is replaced by the local height of the membrane. This
approach has been adopted by several authors considering spatially
inhomogeneous scenarios \cite{bruinsma,barziv,space} and agrees with
scaling arguments based on exact solutions \cite{roland}. For
temperatures nearing the unbinding temperature, where the thermal
fluctuations of the membrane go critical \cite{lipleib}, the validity of
(\ref{helf}) can be questioned and so throughout this paper we will only
consider temperature regimes far from the unbinding transition.
Consequently, for a rough surface, we choose a fluctuation repulsion
satisfying $V_{\rm fluc}(h;z_\rw) = V_{\rm fluc}(h-z_\rw)$, with $V_{\rm
fluc}(h)$ given by (\ref{helf}). 

\subsection{Strongly adhering membranes}

As the membrane now lies much closer to the substrate (a typical distance
can be 30\AA\ as compared with 340\AA\ for the weakly adhering case), the
repulsive interaction becomes dominated by hydration forces. While their
exact origin is not entirely understood, see \cite{isr}, experimentally
they decay exponentially \cite{pars} and have an Angstrom range,
\equn{hyd}
V_{\rm hyd}(h) = b \Si \, \e^{-\alpha h}
\nuqe
where $b$ is a dimensionless number and $\alpha$ an inverse length,
typically $\alpha^{-1} \simeq 2$--$3 \, {\mbox \AA}$. For a
rough substrate we again assume that we can replace $h$ by the relative
height and so use the local expression for the interaction potential,
$V_{\rm hyd}(h-z_\rw)$, in the free energy. Such an
assumption seems reasonable due to the very short-range nature of the
interaction in (\ref{hyd}). 

\subsection{The adhesion energy}

To summarize, the total free energy consists of
\equn{justhereagain}
{\cal F}[h] = \int d\y \Biggl \{ \sqrt{g} \left[ \Si + \half \ka (2H)^2
\right] + p \cdot (h-z_\rw) + V_{\rm vdw}(h,\delta; z_\rw) + V_{\rm
rep}(h-z_\rw)  \Biggr \}
\nuqe
that is, a bending energy, made up of tension (coupled to the membrane
area) and rigidity (coupled to its total mean curvature squared, $H^2$)
terms, and an interaction potential $V$ given by (\ref{nochein}). The
latter contains a contribution from an external pressure, the full
non-local Van der Waals interaction and a local repulsion, $V_{\rm rep}$,
which we consider either as a fluctuation, see (\ref{helf}), or a
hydration interaction as in (\ref{hyd}). 

One of the most relevant quantities in experiments is the membrane
adhesion energy. Within our general mean-field approach, the optimal
height of the membrane is that which minimizes (\ref{justhereagain}). The
value of the free energy of the system when the membrane takes up this
optimum configuration, ${\cal F}_{\rm min}$, leads to a natural
definition of the adhesion energy per unit area
\equn{ad} 
U \equiv - \left( \frac{{\cal F}_{\rm min}}{A_0} - \Si \right)
\nuqe 
Here, $A_0$ is the total area of the {\em projected} 
reference $\y$-plane, $A_0 = \int d\y$. 

Notice that in the above equation a tension dependent term is
subtracted from the definition of the adhesion energy. Doing so
conveniently shifts the origin in such a way that the adhesion energy for
a completely flat membrane has no tension dependent contribution. This
agrees with one's intuitive picture and is necessary as the membrane
tension couples to the entire membrane area and not just to any
excess area arising from a non-planar configuration. Consequently, for a
membrane adhering to a flat wall and being itself flat, $h(\y)=h_0$ for
some constant $h_0$, the adhesion energy is simply given as the negative
of the interaction potential experienced by the membrane. For example,
from (\ref{justhereagain}), [see also (\ref{f})],
\equ
{\cal F}_{\rm min} = A_0 \left[ \Si + V(h_0,\delta;0) \right]
\uqe
and so (\ref{ad}) implies that
\equ
U = - V(h_0,\delta;0)
\uqe
and the tension contribution vanishes. Hence, $U$ is positive for
all sufficiently attractive potentials, $V$. 

\subsection{Rescaling of lengths and interactions}
\label{rescaling}

Before we proceed and calculate the adhesion energy $U$ for different
types of corrugated and rough surfaces, it proves profitable to
extract two natural lengthscales present in the problem.
 
The first of these is provided by the ratio between the Hamaker constant,
$A$, appearing in the Van der Waals potential (\ref{pot}), and the
tension $\Si$ \cite{pgg},
\equ 
a = \left( \frac{A}{2\pi \Si} \right)^\half
\uqe
while the second describes the crossover between the tension and the
rigidity dominated regimes
\equn{xidefn}
\xi = \sqrt{\frac{\ka}{\Si}}
\nuqe

Using these definitions (\ref{f}) becomes
\equn{f2}
\frac{1}{\Si}{\cal F}[h] =\int d\y \Biggl\{ \sqrt{1+(\nabla h)^2}
\left[ 1 + \half \xi^2 \left( \bosy{\nabla} \cdot \frac{\bosy{\nabla}
h}{\sqrt{1+(\nabla h)^2}} \right)^2 \right] +
\frac{1}{\Si}V(h,\delta;z_\rw) \Biggr\}
\nuqe
Equation (\ref{pot}) gives
\equn{wdefn}
\frac{1}{\Si}{W(h;z_\rw)} = \frac{2a^2}{\pi} \int_{h(\y)}^\infty dz \int
d\y' \int_{-\infty}^{z_\rw(\y+\y')} dz' \left[ \y^{'2} + (z-z')^2
\right]^{-(m+1)}
\nuqe
while the fluctuation and hydration interactions can be written,
respectively, as
\equn{Rhelf}
\frac{1}{\Si}V_{\rm fluc}(h) = \frac{3T}{4\pi\ka} {\rm sinh}^{-2}
(\Omega h) 
\nuqe
and
\equn{Rhyd}
\frac{1}{\Si}V_{\rm hyd}(h) = b \, \e^{-\alpha h} 
\nuqe
where $\Omega = ({\pi\Si}/{2T})^{1/2}$ was defined earlier.

In the next sections (\ref{f2}) will be minimized and the adhesion energy
$U$ will be calculated for various corrugated and rough surfaces.
However, first we discuss the case of a simple, flat surface.

\section{A planar surface: choice of physical\\ parameters}
\label{aplanar}
\setcounter{equation}{0}

For a planar and homogeneous solid surface, mean-field theory predicts
that the membrane also adopts a flat configuration,
\equ
h(\y) = h_0
\uqe
where we use the subscript zero to denote all planar quantities.  Here,
we set $z_\rw(\y)=0$ in order to ensure a zero average surface height.
The Van der Waals term (for $m=2$) also simplifies considerably
\equn{plavdw}
\frac{1}{\Si}V_{\rm vdw}(h,\delta;0) = 
-\frac{a^2}{6} \left(
\frac{1}{h^2}-\frac{1}{(h+\delta)^2} \right) 
\nuqe
while the repulsion satisfies (\ref{Rhelf}) and (\ref{Rhyd}) for weak
and strong adhesion, respectively.

The membrane height $h_0$ obeys $V'=\frac{\partial V}{\partial h} = 0$
(balance of forces), that is
\equn{pladiff}
V'(h_0)=
p + \frac{\Si a^2}{3} \left( \frac{1}{h^3_0} - \frac{1}{(h_0 +\delta)^3}
\right) +V'_{\rm rep}(h_0) = 0
\nuqe
with an adhesion energy, (\ref{ad}), of
\eqan{upi}
-\frac{U_0}{\Si} & \equiv & \frac{V_0(h_0,\delta)}{\Si} \nonu \\
&=& \left\{ \begin{array}{ll}
 \frac{p}{\Si} h_0 - \frac{a^2}{6} \left[
\frac{1}{h^2_0}-\frac{1}{(h_0+\delta)^2} \right] + \frac{3T}{4\pi \ka}
{\rm sinh}^{-2} \left( \Omega h_0 \right) & \hspace*{2mm} 
\mbox{weak adhesion} \\
\frac{p}{\Si} h_0 - \frac{a^2}{6} \left[
\frac{1}{h^2_0}-\frac{1}{(h_0+\delta)^2} \right] + b\, \e^{-\alpha h_0} &
\hspace*{2mm} \mbox{strong adhesion}
\end{array} \right.
\naqe
where $V_0(h,\delta) \equiv V(h,\delta;0)$. Equation (\ref{upi}) is
useful as it serves as the starting point to which all our perturbation
theories provide corrections. 

At this point in order to facilitate comparison with experimental systems
we discuss the numerical values of our model parameters. A particular
example is detailed in Table \ref{tab}, where $T = 4.1 \times 10^{-21} \; 
{\rm J}$ at room temperature and $k_B$ was set to unity. For simplicity,
the external pressure is set equal to zero throughout
\begin{equation}
p=0
\end{equation}

\begin{figure}
\begin{center}
\scalebox{0.6}[0.5]{\includegraphics{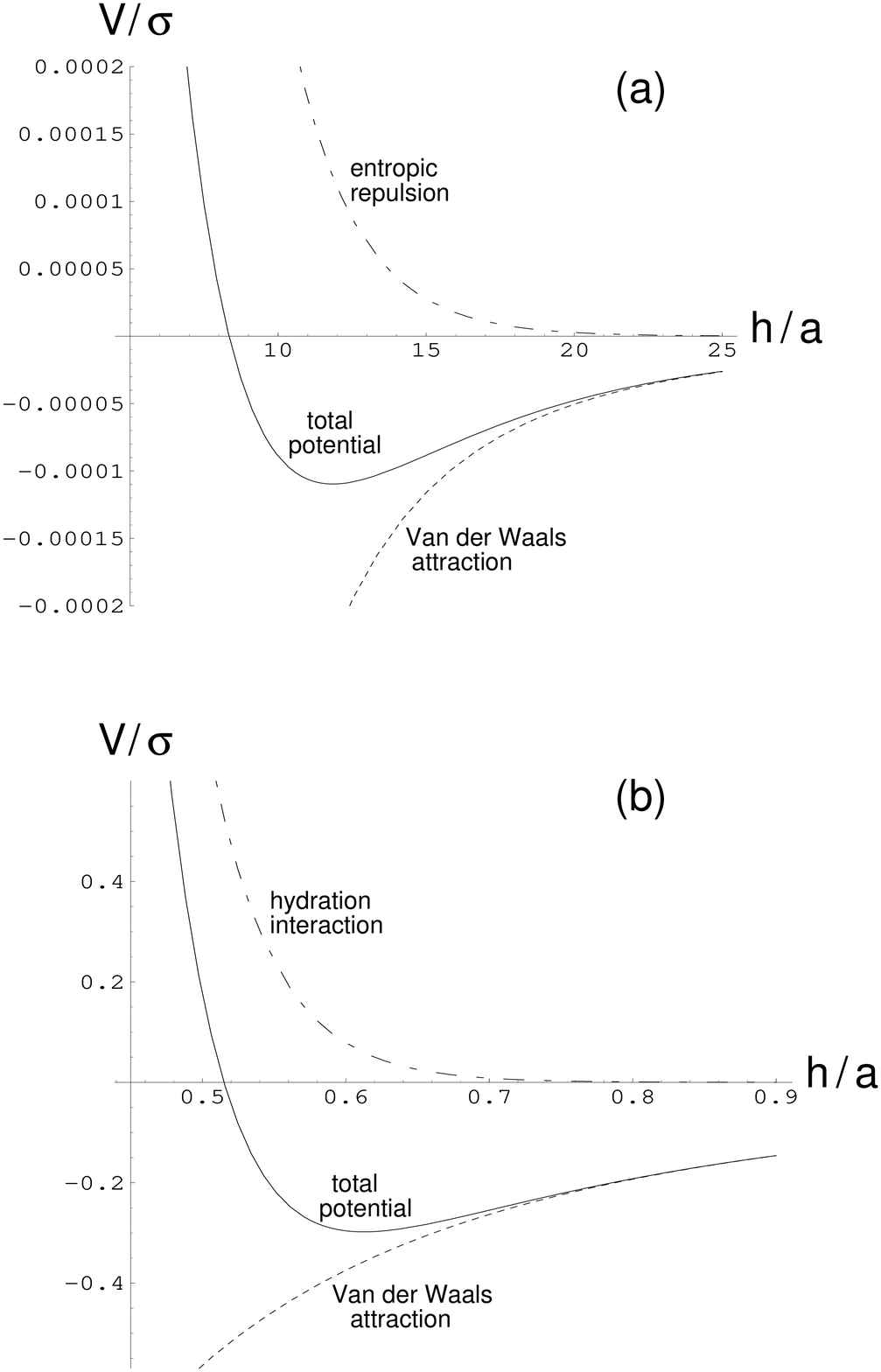}}
\end{center}
\caption{A plot of the various interactions described in Sec.\
\ref{thefree}, with parameter values given in Table \ref{tab}, for a
membrane weakly (a) and strongly (b) adhering to a planar substrate. Here
all potentials are measured in units of the tension, $\sigma = 1.7 \times
10^{-5} {\rm Jm}^{ -2}$, and lengths in terms of $a$. The system can be
seen to equilibrate at $h_0 \simeq 12 a$ and $h_0 \simeq 0.6a$ for weak
and strong adhesion, respectively.}
\label{fig2}
\end{figure}

Let us discuss now the choice of parameters for weak and strong adhering
membranes, separately:

(i)~For weak adhesion, an effective value of the Hamaker constant is
used, $A = 8.67 \times 10^{-22} \, {\rm J} \simeq 0.21\, T$. This value
is quite small in order to approximately model the screening effect of
ions in the solution surrounding the membrane \cite{isr}. Setting $\Si = 1.7 \times 10^{-5 } \; {\rm Jm^{-2}}$, the rescaling
length, $a \simeq 28.5 \; {\mbox \AA}$, and $h_0$ is calculated to be $h_0
\simeq 11.85a\simeq 338 \; {\mbox \AA}$, in agreement with experiment
\cite{rad}. Such a value is reassuring as the expression, (\ref{plavdw}) 
for the Van der Waals interaction, is only valid for $h < 500 \; {\mbox
\AA}$ before retardation effects begin to become important
\cite{lipleib}. From Fig.\ \ref{fig2}, one can see that $V_0(h_0,\delta) 
\simeq -1.1 \times 10^{-4}\Si$ and so the adhesion energy
\eqan{Uwea}
U_0 & \equiv & -V_0(h_0,\delta) \nonu \\
& \simeq & 1.1 \times 10^{-4}\Si = 1.87 \times 10^{-9} \; {\rm Jm^{-2}}
\naqe
which is positive as expected.

(ii)~If we now turn to strong adhesion, ion effects can be ignored and we
use a larger value of the Hamaker constant, $A = 2.6 \times 10^{-21} \;
{\rm Jm^{-2}} \simeq 0.63\,T$ \cite{rad}. This implies that $a \simeq
49.3 \; {\mbox \AA}$ and $h_0 \simeq 0.61a\simeq 30 \, {\mbox \AA}$ in
agreement with measured values using specular reflection of neutrons
\cite{shirl}. The various parameters specifying the hydration force, see
(\ref{hyd}), are
\equ
\begin{array}{lcr}
b = (0.93 \; {\rm Jm^{-2}})/\Si \simeq 5.47 \times 10^4 &;& \alpha^{-1} =
2.2 \; {\mbox \AA}
\end{array}
\uqe
which are in accordance with those measured in \cite{pars}. This time,
see Fig.\ \ref{fig2}, $V_0(h_0,\delta) \simeq -0.298 \Si$ and so
\equn{Ustr}
U_0 \simeq 0.298 \Si = 5.07 \times 10^{-6} \; {\rm Jm^{-2}}
\nuqe
which is significantly greater than (\ref{Uwea}).

Throughout the paper, we will keep to the particular values of the
membrane and external parameters specified here and in Table \ref{tab}. 
It is sometimes convenient to express lengths in terms of the length $a$
and energies (per unit area) in terms of the tension $\sigma$, which is
arbitrarily taken to have the same numerical value for weak and strong
adhering membranes. We should say though that our model can offer only
qualitative, or at best semi-quantitative comparison with experiment.


\section{The Deryagin approximation}
\label{dery}
\setcounter{equation}{0}

One of the most useful approaches, which provides good opportunity for
analytic progress, is the Deryagin approximation \cite{dery1}. First of
all, the full non-local Van der Waals potential, (\ref{pot0}) and
(\ref{pot}), is replaced by a planar potential which is simply a function
of the local relative height coordinate. Since the pressure term is
always local and $V_{\rm rep}$ already has this form,
the total potential is written as
\eqa
V(h,\delta; z_\rw) &\simeq& V_0(h-z_\rw,\delta) \nonu \\
&=& p \cdot (h-z_\rw)
+V_{\rm vdw}(h-z_\rw,\delta;0)+V_{\rm rep}(h-z_\rw)
\aqe
where $V_0(h,\delta) = V(h,\delta;0)$ as before.  Notice that the above
expression corresponds to replacing $z_\rw(\y+\y')$ in (\ref{pot}) with
$z_\rw(\y)$, i.e.\ removing the functional dependence of the potential on
$z_\rw$. The resulting free energy is then expanded to second-order in
$h-h_0$ and $z_\rw$. Since the equilibrium position of the membrane is
given by setting the variation to zero, the first-order term in $h$ and
$z_\rw$ vanishes, yielding
\equn{fdery}
\frac{1}{\Si} {\cal F}[h] \simeq \int d\y \Biggl\{ 1 + \half ({\bf \nabla}
h)^2 + \half \xi^2 ( {\bf \nabla}^2 h)^2 + \frac{1}{\Si}V_0(h_0,\delta) +
\frac{v}{2\Si} (h-h_0-z_\rw)^2 \Biggr\}
\nuqe
with 
\eqan{barv}
v &=& \left. \frac{\partial^2}{\partial h^2} V(h,\delta;0)
\right|_{h=h_0} \nonu \\ 
&=& - a^2 \Si \left[ \frac{1}{h^4_0} -
\frac{1}{(h_0+\delta)^4} \right] + V''_{\rm rep}(h_0) 
\naqe
and
\equ
V''_{\rm rep}(h_0) = \left\{ \begin{array}{ll} 
\frac{3\Si}{4\xi^2} \left[ 2 + \cosh (2 \Omega h_0)
\right] {\rm sinh}^{-4} (\Omega h_0) & \hspace*{1cm} \mbox{weak adhesion} \\
\Si \alpha^2 b \, \e^{-\alpha h_0} & \hspace*{1cm} \mbox{strong adhesion}
\end{array} \right.
\uqe

Writing $\dH(\y) = h(\y) - h_0$, the Euler-Lagrange equation, giving the
value of $\delta h$ which minimizes (\ref{fdery}), is relatively 
straightforward
\equn{el1}
\left( \ka \n^4 - \Si \n^2 + v \right) \dH(\y) = v z_\rw(\y)
\nuqe

To solve (\ref{el1}), it is convenient to convert to Fourier space.
Defining for any function $f(\y)$
\equ
\tilde{f}(\q) = \int d\y f(\y) \e^{-i \q.\y}
\uqe
we find
\equn{fssol}
\tilde{\dH}({\bf q}) = \frac{ \tilde{z}_\rw(\q)}{1+ q^2 \xi_\Si^2+ q^4
\xi_\ka^4}
\nuqe
with the correlation lengths, having the usual definitions \cite{lip}
\equn{corrs}
\begin{array}{lcl}
\xi_\Si^2 = \Si/v &;& \xi_\ka^4 = \kappa /v
\end{array}
\nuqe
For small $q \xi_\Si$, (\ref{fssol}) implies that the membrane follows
the rough surface. However, as $q \xi_\Si$ increases, both the tension
and the rigidity act to dampen this effect. 

Using (\ref{ad}, \ref{fdery}) and (\ref{fssol}), one can write down the
adhesion energy
\eqa
\lefteqn{\frac{U}{\Si} - 1 +\frac{1}{A_0} \int d\y \left\{1+
\frac{V_0(h_0,\delta)}{\Si} \right\} } \nonu \\ 
&= & -\frac{v}{A_0 \Si} \int \frac{d\q}{2(2\pi)^2} \Biggl\{ \left(1+ q^2
\xi^2_\Si + q^4 \xi_\ka^4 \right) | \tilde{\dH}(\q) |^2 +
|\tilde{z}_\rw(\q)|^2 - 2 \tilde{\dH}(\q)  \tilde{z}_\rw(-\q) \Biggr\}
\nonu \\
& = & -\frac{v}{A_0 \Si} \int \frac{d\q}{2(2\pi)^2} \left\{ \frac{ q^2
\xi^2_\Si+ q^4 \xi_\ka^4}{1 + q^2 \xi_\Si^2 + q^4 \xi_\ka^4} \right\}
|\tilde{z}_\rw(\q)|^2
\aqe
By defining the excess in the adhesion energy to be with respect
to the planar case [recall that $U_0=-V_0(h_0,\delta)$]
\eqa
\Delta U & \equiv &  U-U_0 \nonu \\
&=&  U+V_0(h_0,\delta)
\aqe
we obtain 
\equn{Udery}
\frac{\Delta U}{\Si} = - \frac{1}{A_0} \int
\frac{d\q}{2(2\pi \xi_\Si)^2} \left\{ \frac{q^2 \xi^2_\Si + q^4 \xi_\ka^2
}{1 + q^2 \xi_\Si^2 + q^4 \xi_\ka^4} \right\}|\tilde{z}_\rw(\q)|^2
\nuqe
which is only strictly valid up to $O(q^4)$ as higher order terms
in $q$ have not been included in our starting equation.

Equation (\ref{Udery}) has buried within it several assumptions
which we would now like to make more explicit. It is important to realize
that (\ref{fssol}) is not necessarily the general solution to (\ref{el1})
but just a particular solution. The general  solution itself can be found
by adding (\ref{fssol}) to the homogeneous solution, that is the solution
of (\ref{el1}) when $z_\rw$ is set to zero. This then contains the four
constants of integration required to satisfy any boundary conditions.
Therefore , (\ref{Udery}) is only correct when the homogeneous solution
of (\ref{el1}) is identically zero, which is true, for example, at a
sinusoidally corrugated substrate (see Sec.\ \ref{sinusoidally}).
Otherwise, while (\ref{Udery}) is certainly included in $U$, it is not the
whole picture and in this case it may well be better to work entirely in
real space.

For the particular experimental system specified in Sec.\ \ref{aplanar},
one can calculate the various correlation lengths. These have been
included in Table \ref{tab} for completeness.

The approximations used in this section have, as their basis, essentially a
perturbation theory (assuming $h-h_0-z_\rw$ is small) around the planar
value of the adhesion energy, $U_0$. Generally, one can only believe such
an approach if the correction term, $\Delta U$, is much smaller than the
result it is trying to improve upon. Consequently, our findings are
strictly only valid when
\equn{perturb}
\left| \frac{\Delta U}{U_0} \right| \ll 1
\nuqe
which in fact limits the roughness of the substrates we can consider.

\subsection{Sinusoidally corrugated surface}
\label{sinusoidally}

The simplest case to look at is a sinusoidally corrugated surface
\equn{cor}
z_\rw(\y) = c \sin (q x)
\nuqe
with an amplitude $c$ and period $2 \pi/q$. Solving (\ref{el1}),
one can show that
\equn{dhsin}
\dH(\y) = \frac{ c \sin(q x)}{1+q^2 \xi^2_\Si + q^4 \xi_\ka^4} 
\nuqe
Similarly, $\Delta U$ in (\ref{Udery}) is
\equn{Usinusoid}
\frac{\Delta U}{\Si} = - \frac{c^2}{4 \xi^2_\Si} \left(\frac{q^2
\xi_\Si^2 + q^4 \xi_\ka^4}{1+q^2 \xi^2_\Si + q^4 \xi_\ka^4} \right) 
\nuqe
which is valid for small $cq$.

\begin{figure}
\begin{center}
\scalebox{0.6}[0.5]{\includegraphics{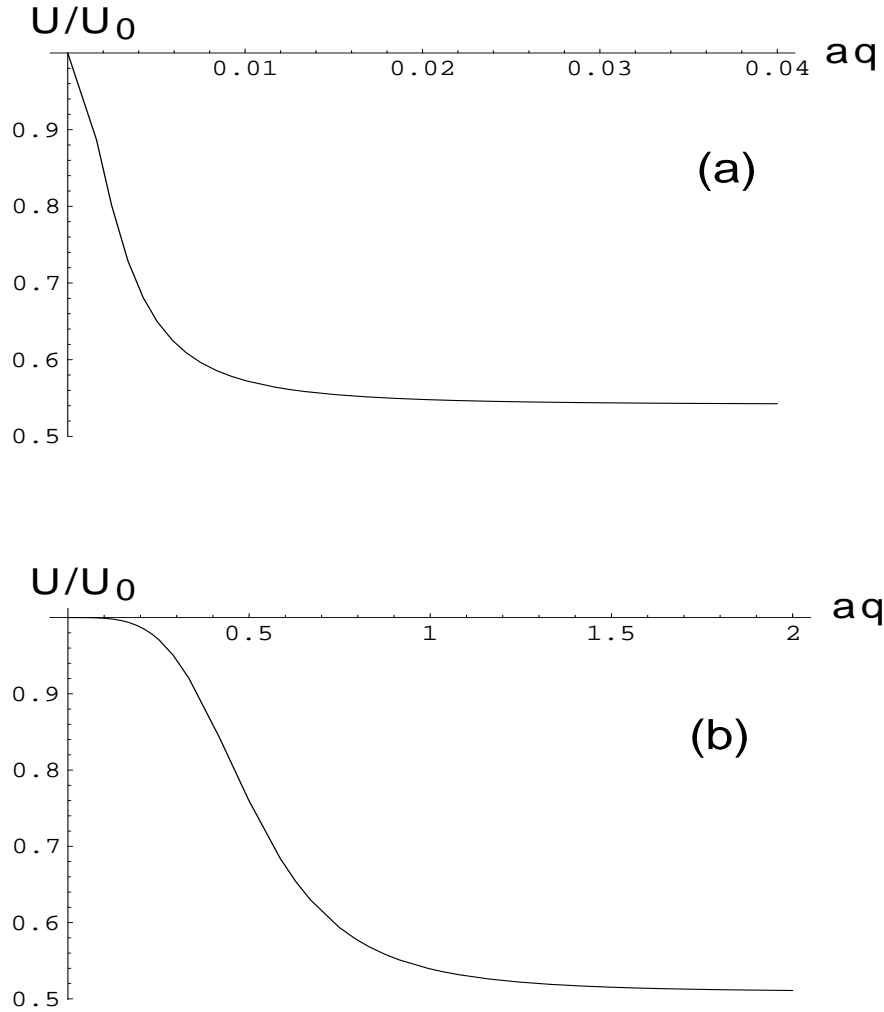}}
\end{center}
\caption{A plot of the adhesion energy versus the rescaled wavenumber $aq$
of a sinusoidally corrugated surface. For weak adhesion (a), the surface
amplitude is set to $c = 5a \simeq 143 \; {\mbox \AA}$ while for a strongly
adhering membrane (b) it is much smaller, $ c = 0.16a \simeq 8 \, {\mbox
\AA}$. All other parameters are given in Table \ref{tab}. For these
values of $c$, $U$ soon reaches a small ($<1$) asymptotic value which
implies that our perturbation theory remains valid for all $q$.}
\label{fig3}
\end{figure}

In Fig.\ \ref{fig3}, $U/U_0$ (using the parameters of Table \ref{tab}) is
plotted as a function of the rescaled wavenumber $aq$ for both weak and
strong adhesion. From (\ref{dhsin}), the average height of the membrane
is unchanged from the planar result $h_0$.  However, as $q$ increases and
the surface becomes corrugated with shorter wavelengths, the adhesion
energy decreases (relative to the planar case) due to the extra bending
energy cost as the membrane tries to follow these surface configurations.
For large $q$, $U$ flattens out; any point on the membrane is almost
equidistant from a crest or trough on the substrate surface and so there
is little energetic benefit in mimicking them. From (\ref{Usinusoid}), we
have, in the limit $q \gg \max( \xi_\sigma^{-1}, \xi_\kappa^{-1})$
\cite{warning},
\equ
\Delta U \simeq - \frac{c^2 \Si}{4\xi_\sigma^2}
\label{lqxi}
\uqe
which is in accordance with $U/U_0 \simeq 0.54$ and $U/U_0 \simeq 0.51$,
see Fig.\ \ref{fig3}, for weak and strong adhesion, respectively. For
small $q \ll \min( \xi_\sigma^{-1}, \xi_\kappa^{-1})$, the membrane can
always follow these long wavelength perturbations,
\equ
\Delta U \simeq -\frac{\Si c^2 q^2}{4}
\uqe
and $U$ remains $q$ dependent.

It is worth pointing out the very different $x$-axis scales and choices
of $c$ in Fig.\ \ref{fig3}. For weak adhesion $c=5a$, that is $c \simeq
143 \; {\mbox \AA}$, the adhesion energy ``bottoms out'' for $q \simeq
0.03/a$. For strong adhesion $c$ is much smaller, $c \simeq 8 \, {\mbox
\AA}$ and much larger values of $q$ are needed before the structure of
the substrate is effectively smeared away. The values of $c$ in both
cases were chosen to ensure that (\ref{perturb}) remains valid and one
can see that $U/U_0$ is never much smaller than $0.5$. Our perturbation
theory seems to work for amplitudes, $c$, which do not grow significantly
bigger than around one third of the height taken by the membrane above a
planar substrate, i.e.\ $h_0/3$. 

\begin{figure}
\begin{center}
\scalebox{0.6}[0.5]{\includegraphics{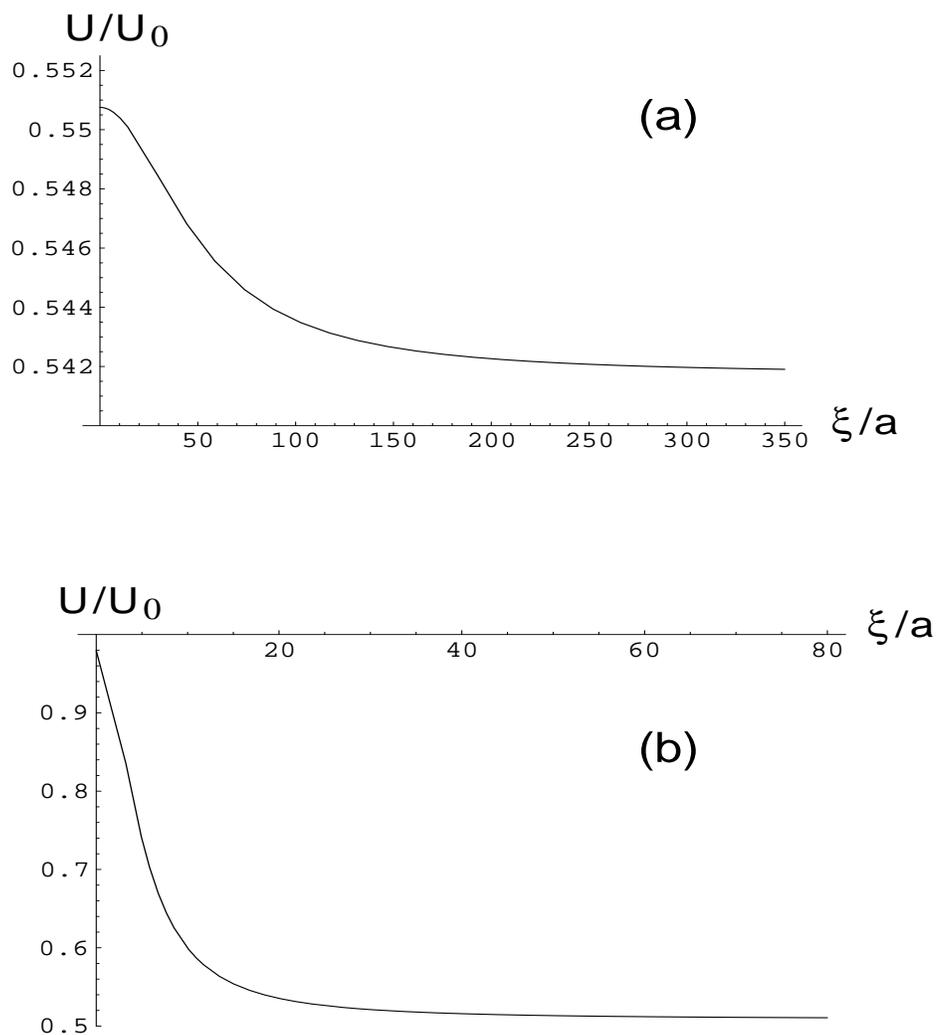}}
\end{center}
\caption{The adhesion energy for a weakly (a) and strongly (b) adhering
membrane above a corrugated surface. This has $c=5a\simeq 143$\AA\ and
$q=0.02/a$ for the former and $c=0.16a\simeq 8$\AA\ and $q=1/a$ for the
latter. Increasing $\xi/a$ corresponds to increasing $\ka/A$. All other
parameter values are given in Table \ref{tab}.}
\label{fig4}
\end{figure}

By noting that $\xi_\ka^2 = \xi \xi_\Si$, (\ref{Usinusoid}) can be
written as
\equn{Usinusoid2}
\frac{\Delta U}{\Si} = -\frac{c^2}{4} \left( \frac{q^2}{\frac{1}{1+q^2
\xi^2} + q^2 \xi_\Si^2} \right) 
\nuqe
and so one can also look at the effect of the elastic modulus $\kappa$ on
the adhesion energy by plotting $U/U_0$ against the crossover length
$\xi/a$ while keeping $\xi_\sigma$ constant, see Fig.\ \ref{fig4}. As
$\xi$ increases, $\ka$ becomes greater than the strength of the Van der
Waals interaction, the Hamaker constant, $A$. This increased bending
energy (coupled with the fact that the membrane average position is
fixed at $h_0$) leads to a raise in magnitude of $\Delta U$ and so the
adhesion energy decreases relative to $U_0$.

\subsection{The profile equation for piecewise
constant surfaces} 
\label{theprofile}

In some cases, and in particular if $z_\rw$ is piece-wise continuous, it
is more convenient to work directly in real space (as opposed to Fourier
space).  The task is then to solve (\ref{el1}) and use (\ref{fdery}) to
find $U$. Given a piecewise constant surface, (\ref{el1})  becomes
\equn{el2}
\left( \ka \n^4 -\Si \n^2 + v \right) \Bigl( \dH(\y) -  z_\rw \Bigr) = 0 
\label{fthorder}
\nuqe
for a constant $z_\rw$ (but which changes value discontinuously for
different ranges of $\y$).

The operator in (\ref{el2}) can be factorized
\equ
\ka \n^4 - \Si \n^2 + v = \ka \left( \n^2 - \eta_+^2 \right) \left(
\n^2 - \eta_-^2 \right) 
\uqe
where $\eta_\pm$ are complex in general and satisfy
\equn{lams}
\eta_\pm = \xi^{-1} \left[ \frac{1 \pm \sqrt{1-4\chi^4}}{2}
\right]^\half
\nuqe
with
\equn{K}
\chi = \left( \ka v/\Si^2 \right)^\frac{1}{4} = \frac{\xi}{\xi_\ka}
\nuqe
a dimensionless ratio of lengths. For our chosen set of parameters (Table
\ref{tab}), $\chi\simeq 0.302$ and $\chi \simeq 9.44$ for weak and strong
adhesion, respectively. 

Therefore, to solve (\ref{el2}) which is a 4$^{th}$ order differential
equation we can consider, equivalently, two coupled 2$^{nd}$ order
differential equations
\eqa
(\n^2 - \eta_-^2) (\dH(\y)-z_\rw) &=& u_0(\y) \\
(\n^2 - \eta_+^2) u_0(\y) &=& 0
\aqe
for the function $\dH(\y)$ and a second function $u_0(\y)$. However,
because of the underlying linearity of the 2$^{nd}$ order differential
equations, the general solution is simply
\equ
\dH(\y) = c_0 u_0(\y) + c_1 u_1(\y) +z_\rw
\uqe
for constant $c_0$, $c_1$ and $z_\rw$, and where $u_1(\y)$ satisfies
\equ
(\n^2 - \eta_-^2) u_1(\y)=0
\uqe
Thus, we have reduced the 4$^{th}$ order differential equation to two
familiar 2$^{nd}$ order Helmholtz equations. This approach is most useful
for systems with cylindrical or spherical symmetry. We note that for
those with one-dimensional symmetry, $z_\rw(\y)=z_\rw(x)$, it is
straightforward to solve the 4$^{th}$ order differential equation
(\ref{fthorder}) directly. The general solution is (for constant $c_0,
c_1, c_2$ and $c_3$) 
\equn{gst}
\dH(x)= c_0 \e^{\eta_+ x} + c_1 \e^{- \eta_+ x } + 
c_2 \e^{\eta_- x} + c_3 \e^{-\eta_- x } + z_\rw
\nuqe
which we will use in the following sections.

\subsection{An isolated trench}
\label{anisol}

\begin{figure}
\begin{center}
\scalebox{0.58}{\includegraphics{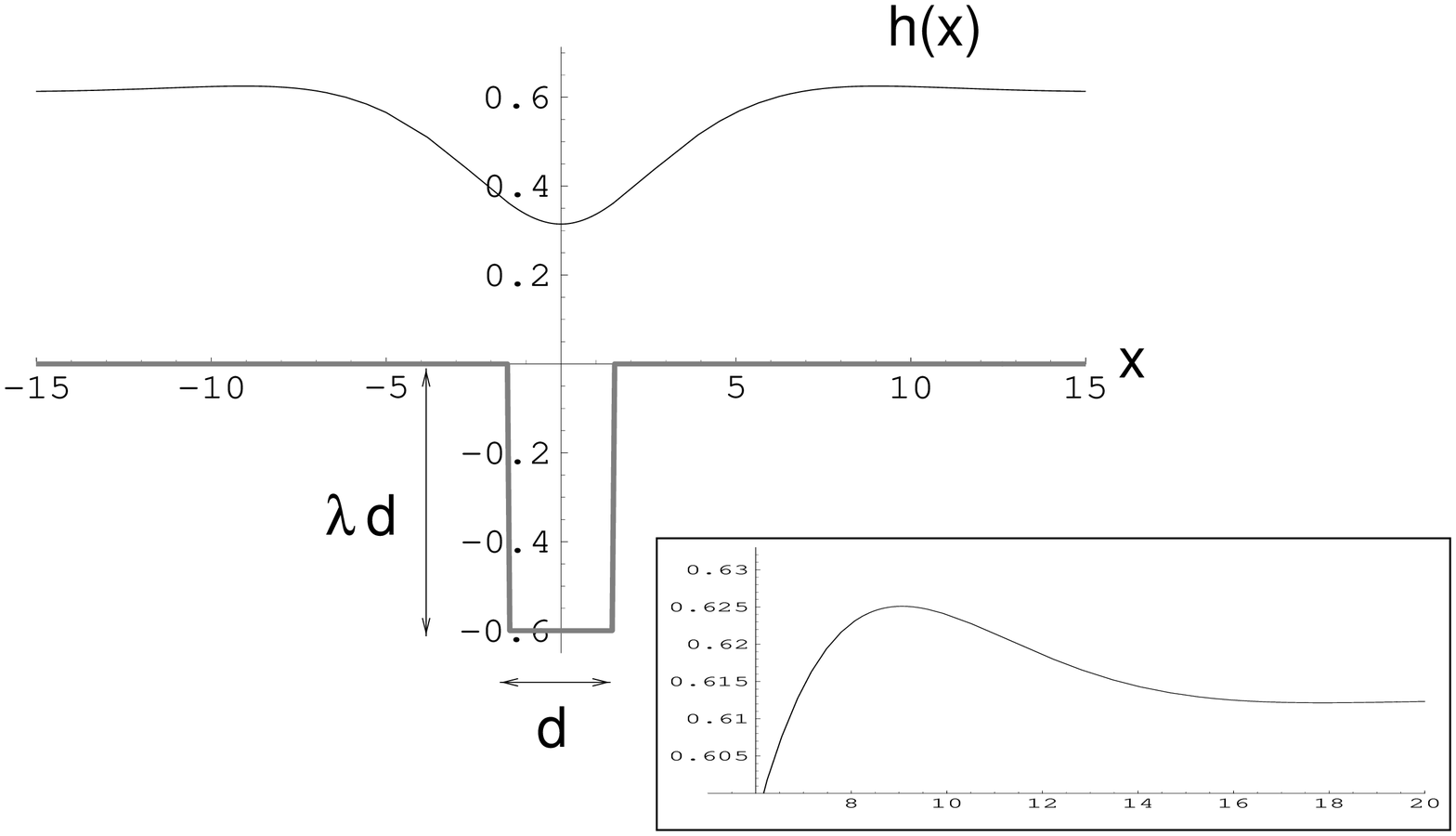}}
\end{center}
\caption{The membrane profile $h(x)$ for strong adhesion to a solid surface
(thick line) broken by a single trench of width $d=3a\simeq 148$\AA\ and
depth $\lambda d$ with $\lambda=0.2$. The inset region shows a blow up of
the profile with scales chosen so as to emphasize the overshoot present. 
The local adhesion energy $U/U_0 \simeq 0.51$. All lengths are stated in
units of $a$ without explicit statements for clarity. Parameter values
are given in Table \ref{tab} (strong adhesion) and only the lower lipid
leaflet is shown.}
\label{fig5}
\end{figure}

We next consider a single, infinitely long trench parallel to the $y$
axis and of width $d$ and depth $\lambda d$. The height profile of the
surface takes either of two fixed values: $-\lambda d$ for $|x|<d/2$, and
zero otherwise (see Fig.\ \ref{fig5}).  This can be also written as
\equ
z_\rw(\y)=z_\rw(x) = -\lambda d \Bigl\{ \theta(x+d/2) - \theta (x-d/2)
\Bigr\}
\uqe
where $\theta(x)$ is the Heaviside step function, being only non-zero
(and equal to unity) for positive $x$. The one-dimensional solution of
(\ref{el2}) is given by (\ref{gst}) with $z_\rw$ taking either of the
fixed values $-\lambda d$ for $|x| < d/2$ or zero otherwise. The
constants of integration can be determined from the following boundary
conditions
\eqa
\dH(-x) &=& \dH(x)  \\
\dH(\pm x) &\goto& 0 \mbox{\hspace*{15mm} as $|x| \goto \infty$}
\aqe
and by imposing continuity (up to the third derivative in $x$) at the
edges of the trench, $x = \pm d/2$.  The final solution is (for positive
$x \ge 0$)
\equn{fsol}
\frac{1}{\lambda d}\dH(x) = \left\{ 
\begin{array}{ll}

\begin{array}{l}
 -1 + \half k_{-}\e^{-\frac{\eta_{+}d}{2}} \cosh (\eta_{+}x) 
\\ \hspace*{6mm} + 
\half k_{+} \e^{-\frac{\eta_{-}d}{2}} \cosh (\eta_{-}x)  
\end{array} & \mbox{\hspace*{8mm} for $0 \le x \le d/2$} \\
\\
\begin{array}{l} 
-\half k_{-} \sinh \left( \frac{\eta_{+}d}{2} \right)\e^{-\eta_{+}x}
\\ 
-\half k_{+} \sinh \left( \frac{\eta_{-}d}{2} \right) \e^{-\eta_{-}x}
\end{array} & \mbox{\hspace*{8mm} for $x \ge d/2$}

\end{array} 
\right.
\nuqe
and
\equn{kpm}
k_\pm=1 \pm \frac{1}{\sqrt{1-4\chi^4}}
\uqe

In Fig.\ \ref{fig5}, the membrane profile is sketched for a supported
membrane (strong adhesion case) with the same parameters as those given
in Table \ref{tab}. The trench is specified by $d=3a$, i.e.\ $148 \, {\mbox
\AA}$, and $\lambda = 0.2$. The membrane can be seen to follow the
contour of the surface, as expected from the general prediction
(\ref{fssol}), without developing a similar discontinuity to that
occurring at the boundary of the trench. In addition, it can be seen that
an overshoot is present, with the membrane having a height greater than
$h_0$ (for an extensive discussion of overshoots see \cite{space}). From
(\ref{fsol}), $\dH=h-h_0$ can be shown to be negative for all $x > d/2$,
providing $\eta_\pm$ are real, and so the overshoot can only arise if
$\eta_\pm$ go complex. That is, from (\ref{lams}), when $1-4\chi^4<0$ or
\equn{overy}
 2 \xi > \xi_\Si \mbox{\hspace*{10mm} necessary condition for overshoot}
\nuqe
i.e.\ in the rigidity dominated regime ($4\kappa > \Si^2/v$).

Equation (\ref{overy}) can be understood by realizing that the rigidity
term in the free energy (\ref{fdery}) prevents the membrane from turning
any sharp corners (see Appendix).  For a pure interface ($\ka=0$ and $\Si
\ne 0$), (\ref{overy}) cannot be satisfied and, within our Deryagin 
approximation,  no overshoot ever occurs. 

\begin{figure}
\begin{center}
\scalebox{0.58}{\includegraphics{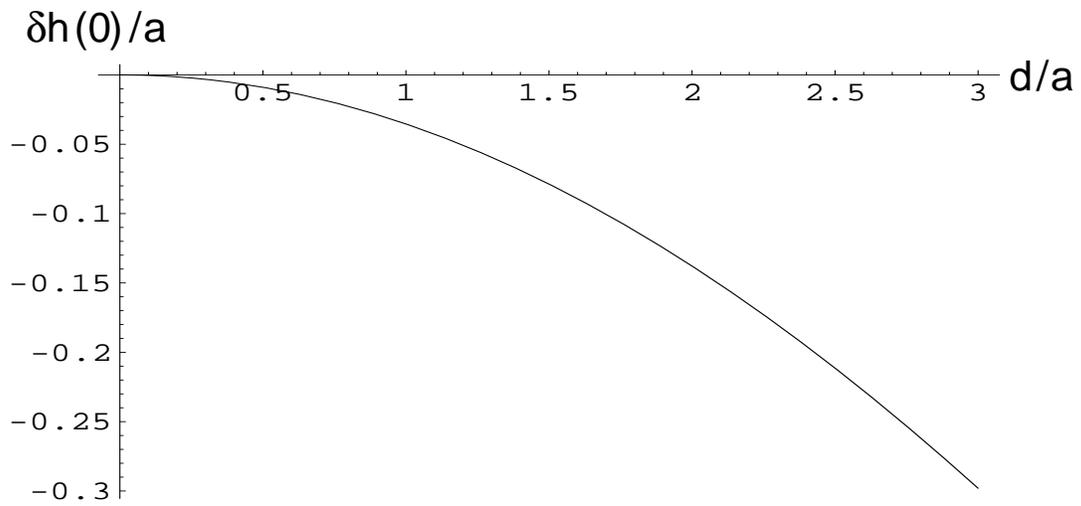}}
\end{center}
\caption{The height of the membrane $\delta h(0)= h(0)-h_0$ at the center
of an isolated trench of width $d$ and depth $\lambda d$, where $\lambda
= 0.2$. See Table \ref{tab} (strong adhesion) for choices of the other
parameters.}
\label{fig6}
\end{figure}

It is also interesting to look at the extent to which the membrane
penetrates into the trench. A natural measure of this quantity is the
membrane height in the trench center: 
\equn{justhere}
\dH(0)=\lambda d \left(-1+\half k_{-}\e^{-\eta_{+}d}
+\half k_{+}\e^{-\eta_{-}d} \right)
\nuqe
In Fig.\ \ref{fig6} this quantity is plotted against $d /a$, the trench
width [here, $\lambda$ simply scales out -- see (\ref{justhere})].  For
larger $d$, the membrane is able to enter the trench more easily as there
is less bending energy cost in the smoother configuration required to do
so.  For small $d/a$, $\dH(0)\goto 0$ and $h(0)$ tends towards the planar
value of the average membrane height as the trench becomes increasingly
more narrow.  In the opposite limit (not yet visible in Fig.\
\ref{fig6}), $d/a \gg 1$, (\ref{justhere}) implies that $\dH(0)/a \goto -
\lambda d/a$, though here, we are pushing our perturbation theory beyond
its region of validity. 

The adhesion energy (\ref{ad}) is defined as an energy per unit of
projected area. In order to obtain a finite contribution in the case of
an isolated surface perturbation like a trench, we need to consider one
embedded in a flat surface of {\it finite} lateral extent $L$.  Hence,
using a local definition, involving a cut-off $L$, we can write that the
change in $U$, $\Delta U$ (up to second-order in $\dH$) as
\equn{tw1}
\Delta U = - \frac{1}{2L} \int_0^L dx \left\{ \Si \bigl( \dH^{'} \bigr)^2
+\ka \bigl( \dH^{''} \bigr)^2 + v \left( \dH - z_\rw \right)^2 \right\}
\nuqe
From (\ref{gst}), we can see, that for real $\eta$, the cut-off should be
proportional to $\eta_+^{-1}$ due to the exponential decay. However, for
$\chi>1/\sqrt{2}$ both $\eta_+$ and $\eta_-$ are no longer real and it is
straightforward to show that
\equn{etacom}
\eta_+ = \eta_-^* = \half \xi^{-1} \left\{ \sqrt{2 \chi^2+1} +
i \sqrt{2 \chi^2 -1} \right\}
\nuqe
where the asterix denotes complex conjugation. In this case, (\ref{gst})
implies that for $x > d/2$ the profile has an exponential term,
$\exp \left( - \frac{\sqrt{2 \chi^2+1}}{2 \xi}x \right)$, and so a
natural choice for $L$ is
\equn{cutoff}
L = 8 \times \max \left( \frac{d}{2} , \frac{1}{{\Re} \{ \eta_+ \}}
\right) 
\nuqe
where the prefactor was chosen by examining some numerical solutions for
the profile. If $\eta_+$ is complex then (\ref{etacom}) implies that
$1/{\Re} \{ \eta_+ \} = \frac{2\xi}{\sqrt{2\chi^2+1}}$. Using this
definition $L \simeq 22.3a $ for the system specified in Table \ref{tab},
which is a sensible value (see Fig.\ \ref{fig5}). 

Using (\ref{fsol}) and (\ref{tw1}), one can show that
\eqan{tw2}
\frac{\Delta U}{\Si} &=& -\frac{\lambda^2 d}{32 L } \Biggl\{
\left( \frac{k_{-}}{\Lambda_{+}}\right)^2 
I(\Lambda_{+},\Lambda_{+}) + 
\left( \frac{k_{+}}{\Lambda_{-}} \right)^2
I(\Lambda_{-},\Lambda_{-}) \nonu \\ 
& & + \left( \frac{2\xi}{d} \right)^2 k_{+} 
k_{-} \, I(\Lambda_{+},\Lambda_{-}) \Biggr\}
\naqe
where $\Lambda_\pm\equiv \eta_\pm d$ is introduced,
\eqa
I(u,v) &=& u v (u+v)(\e^u-1)(\e^v-1) \, \e^{-(u+v)\left( \half+
\frac{L}{d} \right)} \nonu \\
& & \times \left[ \frac{ \e^{(u+v)\left(
\frac{L}{d} -\half\right)}}{1-\frac{u (\e^v-1)+v(\e^u-1)}
{u \e^u(\e^v-1) + v \e^v(\e^u-1)}} - \half \right] 
\aqe
and $k_\pm$ are defined in (\ref{kpm}).  The function $I(u,v)$ can be
seen by inspection to be positive, given that $2 L \ge d$ (from
(\ref{cutoff})), implying that $\Delta U$ is negative. Any non-planar
membrane configuration (and such a configuration must be adopted by a
membrane adhering to a rough substrate) will give an additional (and
dominant) bending energy cost in the definition (\ref{ad}) and so
decrease $U$. 

\begin{figure}
\begin{center}
\scalebox{0.58}{\includegraphics{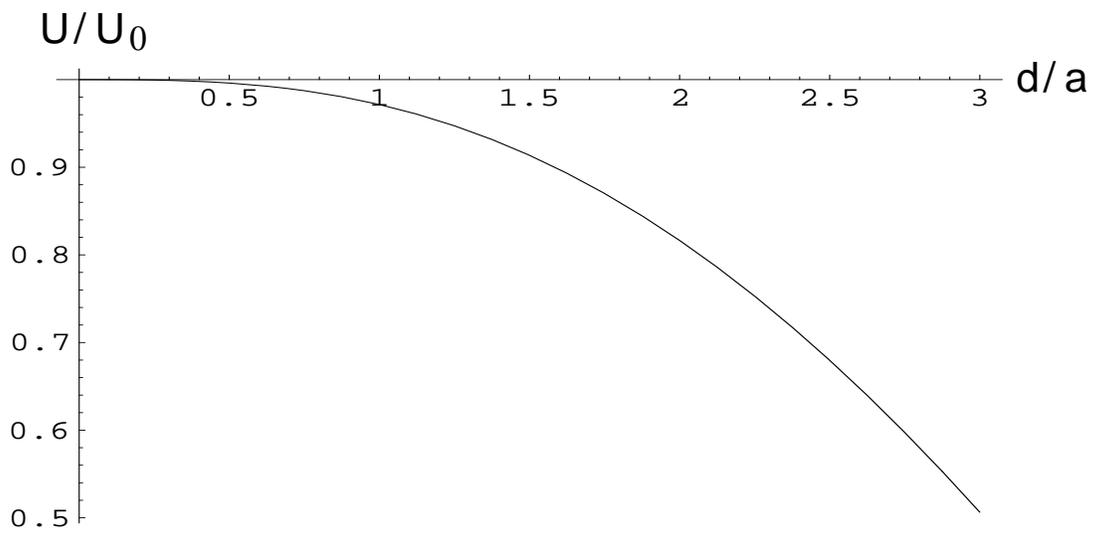}}
\end{center}
\caption{The relative adhesion energy, $U/U_0$ plotted for a trench of
width $d$ and depth $\lambda d$, with $\lambda=0.2$.  For an isolated,
non-planar perturbation of the substrate, a local definition of the
adhesion energy needs to be used and involves the cut-off $L$ (see text). 
All other parameters used here are given in Table \ref{tab} (strong
adhesion case).}
\label{fig7}
\end{figure}

Equation (\ref{tw2}) is illustrated in Fig.\ \ref{fig7}. The adhesion
energy decreases for larger $d$ as the membrane adopts a more and more
non-planar configuration. For small $d/a$, one can show that (\ref{tw2})
becomes
\equ
\frac{\Delta U}{\lambda^2 \Si} = - \frac{a^3 \chi^4}{4 \xi^2 L} \cdot
\left( \frac{d}{a} \right)^3 + O \Bigl(\Bigl(\frac{d}{a}\Bigr)^4 \Bigr) 
\uqe
and vanishes as $d \goto 0$ in agreement with Fig.\ \ref{fig7}. For
larger values of $d/a$, Fig.\ \ref{fig7} would seem to imply that $\Delta
U$ continues to grow in magnitude. This, of course, is false and is an
unfortunate artifact of going beyond the valid limit of our perturbation
theory. Exact numerical solutions for a similar scenario will be
discussed in a companion paper \cite{II} and do not exhibit such
unphysical behavior.

Again, we wish to emphasis that the parameters specifying the substrate
geometry ($\lambda$ and $d$ in this case) were chosen so that the
constraint (\ref{perturb}) was obeyed.

\subsection{An isolated pit}
\label{anisolpit}

We next consider a single cylindrical pit of radius $r$ and depth
$\lambda r$. Equation (\ref{el2}) is now only a function of
$\rho=\sqrt{x^2+y^2}$ and is solved by Bessel functions, as is usual for
systems with cylindrical symmetry,
\equ
\dH(\rho) = c_0 I_0(\rho \eta_-) + c_1 K_0(\rho \eta_-) + c_2
I_0(\rho \eta_+) + c_3 K_0(\rho \eta_+) + z_\rw
\uqe
Here, $z_\rw = - \lambda r$ for $\rho<r$ and is zero elsewhere. Requiring a
finite solution at $\rho=0$ and a vanishing one as $\rho \goto \infty$,
implies that
\equn{pitans}
\dH(\rho) = \left\{ 
\begin{array}{lr} 
c_0 I_0(\rho \eta_-) + c_2 I_0(\rho \eta_+) - \lambda r &
\mbox{\hspace*{5mm} for $\rho<r$} \\ 
c_1 K_0(\rho \eta_-) + c_3 K_0(\rho \eta_+) & \mbox{\hspace*{5mm}
for $\rho>r$}
\end{array} 
\right.
\nuqe
The constants of integration can be found by imposing continuity at
$\rho=r$. 

\begin{figure}
\begin{center}
\scalebox{0.6}{\includegraphics{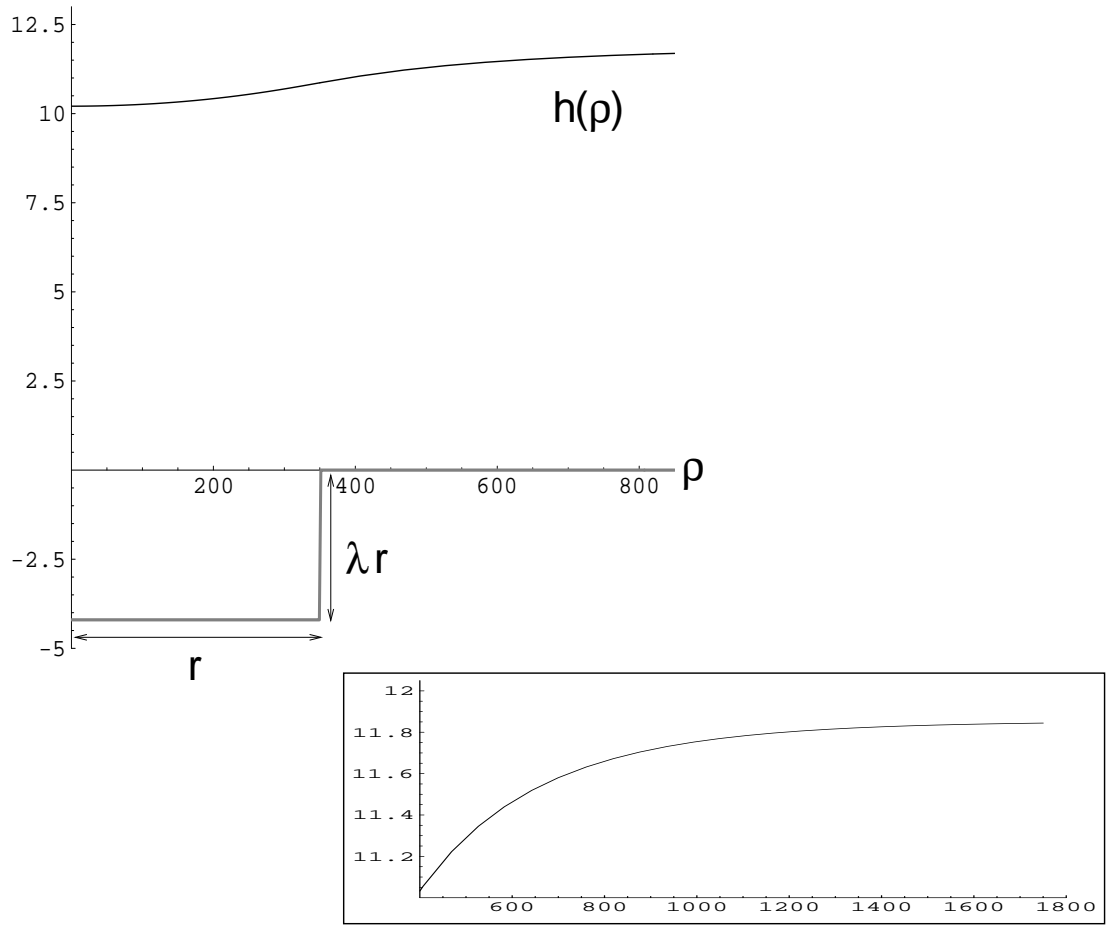}}
\end{center}
\caption{A typical membrane configuration $h(\rho)$ (of the lower lipid
leaflet)  weakly adhering over a cylindrically symmetric pit of depth
$\lambda r$ and radius $r$. Here $\lambda = 0.012$ and $r=350a \simeq
0.997\;\mu$m (for further choices see Table \ref{tab} -- weak adhesion).
The local adhesion energy (with $L \simeq 259.1a$) is $U/U_0 \simeq
0.50$. The inset shows that no overshoot is present as $\eta_\pm$ are now
real. All lengths are given in units of $a$.}
\label{fig8}
\end{figure}

The cylindrical symmetry results in little qualitative changes from the
previous section (see Fig.\ \ref{fig8}). Though, our perturbation theory
is now acceptable up to a large hole of radius $r = 350a \simeq
0.997\, \mu{\rm m}$ due to the much smaller non-planar area of the
substrate. A quick glance at Table \ref{tab} shows that in this case
$\eta_\pm$ are real and from (\ref{lams}) and (\ref{overy}) no overshoot
can then occur \cite{over}. This can also be seen in Fig.\ \ref{fig8}. A
local adhesion energy can be defined similarly to (\ref{tw1}) with the
cut-off $L$ obeying
\equ
L = 8 \times \max \left(r, \frac{1}{ {\Re} \{ \eta_+ \}} \right)
\uqe

Unfortunately, for this case $\Delta U$ is a very complicated expression
and it is not possible to proceed analytically. Therefore, similar to Sec.\
\ref{anisol}, $\Delta U$ is calculated, via (\ref{tw1}), with the
solution (\ref{pitans}) but then the integrals are evaluated numerically.
One finds that $U$ decreases with $r$ in a similar manner to Fig.\
\ref{fig7}. 

\subsection{Periodic array of trenches}
\label{trsSec}

\begin{figure}
\begin{center}
\scalebox{0.6}{\includegraphics{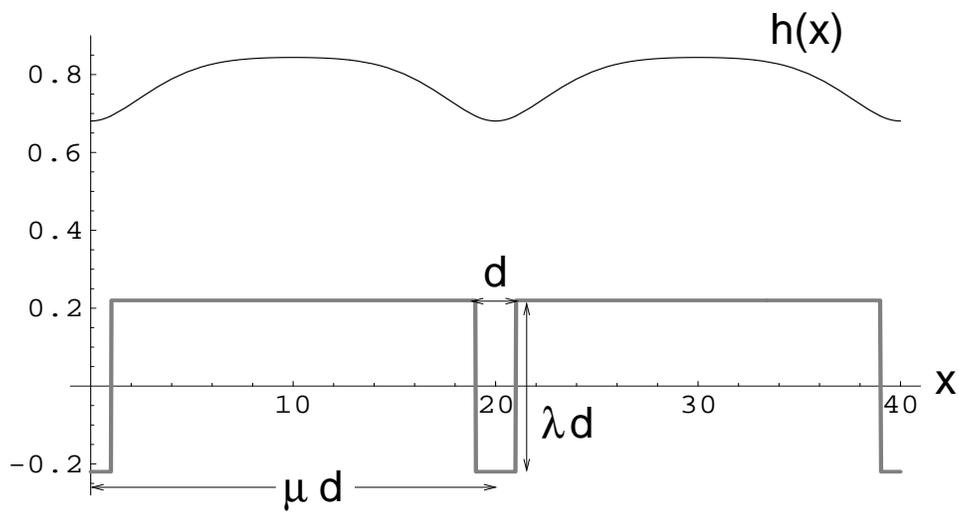}}
\end{center}
\caption{The profile of a membrane $h(x)$ supported (strong adhesion) 
above a periodically structured substrate with trenches of infinite
length, width $d$ and depth $\lambda d$. We have set $\lambda = 0.22$,
$\mu = 10$ and $d=2a \simeq 99 \; {\mbox \AA}$ in the above.  The membrane
configuration (thin line) follows that of the substrate (thick line) but
at a much reduced amplitude (due to the effects of rigidity and tension).
All lengths are shown in units of $a$.}
\label{fig9}
\end{figure}

The next, more complicated, scenario to consider is a periodic array of
one-dimensional trenches. These are of depth $\lambda d$ and width $d$ in
the $x$ direction, and infinitely long in the $y$ direction.  We let the
length of the repeating unit making up the surface be $\mu d$ (see Fig.\
\ref{fig9}), then
\equ
z_\rw(\y) = \half \lambda d - 
\lambda d\sum_{n=-\infty}^{\infty} 
\left[\theta(x-n\mu d + d/2) - \theta(x-n \mu d - d/2) \right]
\uqe
summing over integer $n$.

By ensuring that the solution is locally symmetric about each trench
center, it is possible to write out, using (\ref{gst}), $\dH$ explicitly
over one period
\newcounter{d}
\begin{list}{$\bullet$}{\usecounter{d}}
\item for $0 < x < d/2$ 
\equ
\dH(x) = c_0 \cosh(\eta_+ x) + c_1 \cosh(\eta_- x) - 
\frac{\lambda d}{2}
\uqe
\item for $d/2 < x < \mu d - d/2$
\equ
\dH(x) = c_2 \e^{\eta_+ x} + c_3 \e^{-\eta_+ x} + c_4 
\e^{\eta_- x} + c_5 \e^{-\eta_- x} + \frac{\lambda d}{2}
\uqe
\item for $\mu d - d/2 < x < \mu d$
\equ
\dH(x) = c_{0} \cosh \Bigl(\eta_+(x-\mu d) \Bigr) + 
c_{1} \cosh \Bigl(\eta_-(x -\mu d) \Bigr) - \frac{\lambda d}{2}
\uqe
\end{list}
As the configuration taken up by the membrane is symmetric around $x=\mu
d/2$,
\equ
\dH(x) = \dH(\mu d-x) 
\uqe
one can show that
\equ
\begin{array}{lcl}
c_3=c_2 \e^{\eta_+\mu d} \\
c_5=c_4 \e^{\eta_- \mu d} 
\end{array}
\uqe
and so reduce the number of unknowns to four. These can be found by
imposing continuity of $\dH$ and its derivatives at $x = d/2$.

The change in adhesion energy  (to second-order in $\dH$) is
\eqa
\Delta U &=& - \frac{1}{\mu d}\int_0^{d/2} dx \left[ \Si
\bigl( \dH^{'} \bigr)^2 + \ka \bigl( \dH^{''} \bigr)^2 + v (\dH + \lambda d/2)^2 \right] \nonu \\ 
& &  -\frac{1}{\mu d}\int_{d/2}^{\mu d/2} dx 
\left[ \Si \bigl( \dH^{'} \bigr)^2 + \ka \bigl(\dH^{''} \bigr)^2 + v
(\dH - \lambda d/2)^2 \right]
\aqe
and after some algebra, can be found to satisfy
\equn{treU}
\frac{\Delta U}{\Si} = -\frac{\lambda^2}{2 \mu} \Biggl[
I(\Lambda_{+},\Lambda_{-}) + I(\Lambda_{-},\Lambda_{+}) \Biggr]
\nuqe
where we define
\equn{treU2}
I(u,v) = \frac{(\e^u-1)(\e^{(\mu-1)u}-1)}{\e^{\mu u}-1} \cdot
\frac{u v^2}{(u^2-v^2)^2} \left[ v^2-2 \left( \frac{d}{\xi} \right)^2
\chi^4 \right]
\nuqe
recalling that $\Lambda_\pm = \eta_\pm d$. Again, it is possible to show
that $\Delta U$ is always negative. Note, that the dependence of the
adhesion energy on the depth of the trenches, via $\lambda$, is, from
(\ref{treU}), simply parabolic (providing (\ref{perturb}) holds). 

\begin{figure}
\begin{center}
\scalebox{0.6}{\includegraphics{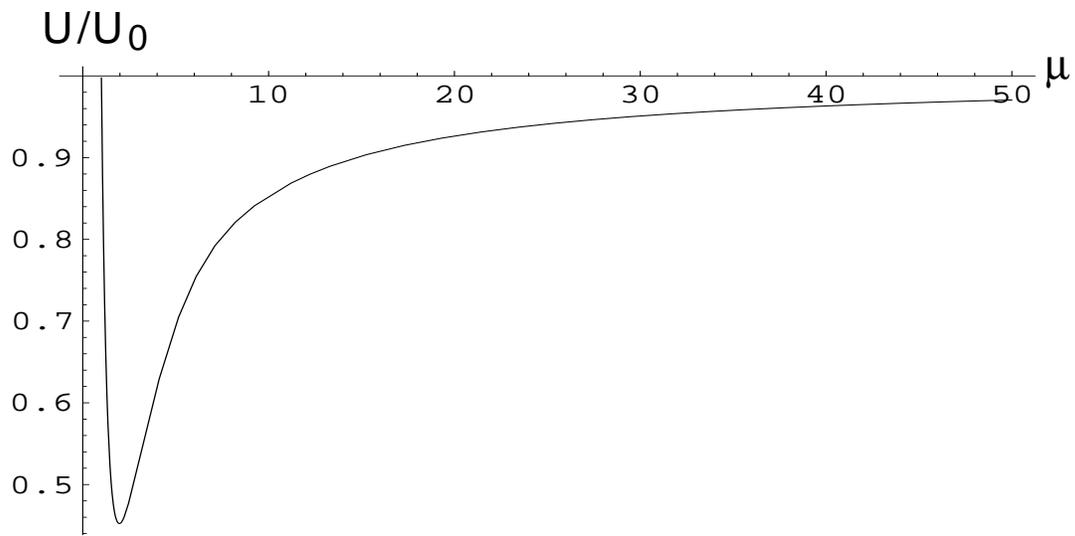}}
\end{center}
\caption{For $\lambda=0.12$ and $d=2a\simeq 99$\AA, the relative adhesion
energy $U/U_0$ in the strong adhesion case is plotted against $\mu$, the
periodicity parameter. For the two flat limiting surfaces, $\mu=1$ and
$\mu=\infty$, the decrement $\Delta U$ is zero and $U=U_0$. The minimum
in the curve occurs near $\mu=2$ as described in Sec.\ \ref{trsSec}.}
\label{fig10}
\end{figure}

For a surface roughness of this form, an interesting consequence of
(\ref{treU}) is that the excess adhesion energy, $\Delta U$, has a
minimum as a function of $\mu$ at $\mu = \mu^*$. This is illustrated for
a strongly adhering membrane in Fig.\ \ref{fig10}. If we use $\epsilon
\equiv d/\xi$ as an expansion parameter, (for Fig.\ \ref{fig10},
$\epsilon \simeq 0.11$), then perturbation theory gives
\equn{uexpan}
\frac{\Delta U}{\lambda^2 \Si} = - \frac{\mu-1}{2 \mu^2} \chi^4
\epsilon^2 + \frac{(\mu-1)^2}{1440\mu} \cdot (\mu^2+2\mu-2) \chi^8
\epsilon^6 + O \Bigl( \epsilon^8 \Bigr) 
\nuqe
which implies
\equn{mu*}
\mu^* = 2 - \frac{\chi^4}{30} \epsilon^4 + O \Bigl( \epsilon^6 \Bigr)
\nuqe

This can be understood by looking at the average height of the surface,
\eqan{avhei}
\langle z_\rw \rangle &=& \frac{1}{\mu d} \int_0^{\mu d} dx \, z_\rw(x)
\nonu \\ 
&=& \frac{\lambda d}{2} \left(1-\frac{2}{\mu} \right) 
\naqe
For $1<\mu<2$, as $\mu$ increases, (\ref{avhei}) evaluated near $\mu=1$,
implies that the surface can be visualized as an array of spikes
perturbing an ``initial" planar state with $\langle z_\rw \rangle =
-\lambda d/2$. For $\mu=1$ the substrate is flat and consequently the
general solution, (\ref{gst}), implies that the membrane is so also, with
$\dH = -\lambda d/2$ and $U=U_0$.  As the thickness of the spikes
increases, the membrane responds and is forced to bend more and more.
This costs energy and $\Delta U$ increases in magnitude.  One can show
that
\equ
\frac{\Delta U}{\Si} = 
- \half{\Bigl(\frac{\lambda d \chi^2}{\xi}\Bigr)}^2 (\mu-1)
+ O \Bigl( (\mu-1)^2 \Bigr) 
\uqe
for $\mu \simeq 1$. However, for $\mu>2$, the surface (and hence the
membrane) starts to tend towards another, ``final", planar state, that of
$\langle z_\rw \rangle = \lambda d/2$ which occurs when $\mu = \infty$. 
As $\dH$ of the membrane, at this extreme point, also satisfies $\dH =
\lambda d/2$, $| \Delta U|$ decreases with growing $\mu$. Looking at the two
extreme values, $\mu=1$ and $\mu=\infty$, one can see, from (\ref{treU})
and (\ref{treU2}), that $\Delta U$ vanishes as expected.

This behavior becomes more transparent if the variance in the height of
the surface is considered
\eqan{vari}
\langle \Delta z_\rw^2 \rangle &=& \langle ( z_\rw - \langle z_\rw
\rangle )^2 \rangle \nonu \\
&=& \left( \frac{\lambda d}{\mu} \right)^2 \cdot (\mu-1)
\naqe
with the average defined in (\ref{avhei}). Equation (\ref{vari}) vanishes
for $\mu=1$ and $\mu=\infty$, and has a maximum at $\mu=2$. Therefore, we
find that the excess adhesion energy is greatest (in magnitude) for that
value of $\mu$ at which the surface is the most rough, i.e.\ $\mu = 2$.
In fact, if $U$ is plotted against $\langle \Delta z_\rw^2 \rangle$, it
is just a monotonically decreasing function. From (\ref{uexpan}) and
(\ref{vari}), we have
\equ
\frac{\Delta U}{\sigma} = - \frac{\chi^4 \langle \Delta z_\rw^2
\rangle}{2d^2} \epsilon^2 + O \Bigl( \epsilon^6 \Bigr) 
\uqe
and, at this level of approximation, there is a simple, linear
relationship. The average height of the membrane itself can also be
calculated
\eqa
\langle h \rangle &=& h_0 + \langle \dH \rangle \nonu \\
&=& h_0+ \frac{\lambda d}{2} \left(1-\frac{2}{\mu} \right) 
\aqe
and $\mu=2$ again arises as a significant value. 

It is also interesting to examine the extent to which the membrane
penetrates each of the trenches.  Concentrating on $\delta h(0)$, we find
\equ
\frac{\delta h(0)}{\lambda d} = \Lambda_{+}^2 \Lambda_{-}^2 \left\{
\frac{I(\Lambda_{+}) - I(\Lambda_{-})}{\Lambda_{+}^2-\Lambda_{-}^2}
\right\} - \half
\uqe
where
\equ
I(u) = - \frac{\e^{-u/2}}{u^2} \cdot \frac{\e^{\mu u}-
\e^u}{\e^{\mu u}-1}
\uqe
and $\Lambda_\pm = \eta_\pm d$. Hence, at the level of the Deryagin 
approximation, there is a linear dependence on the trench depth, 
 $\lambda d$. Looking at $h_0 + \delta
h(0) - \langle z_\rw \rangle$, from (\ref{avhei}), one can see (as $I(u)$
is a monotonically increasing function) that the membrane always
penetrates further into the trenches as their depth grows. It experiences
a more attractive potential due to the greater surface area of the
substrate. However, one must be aware that increasing $\lambda$ can
quickly cause (\ref{perturb}) to be broken.

\begin{figure}
\begin{center}
\scalebox{0.6}{\includegraphics{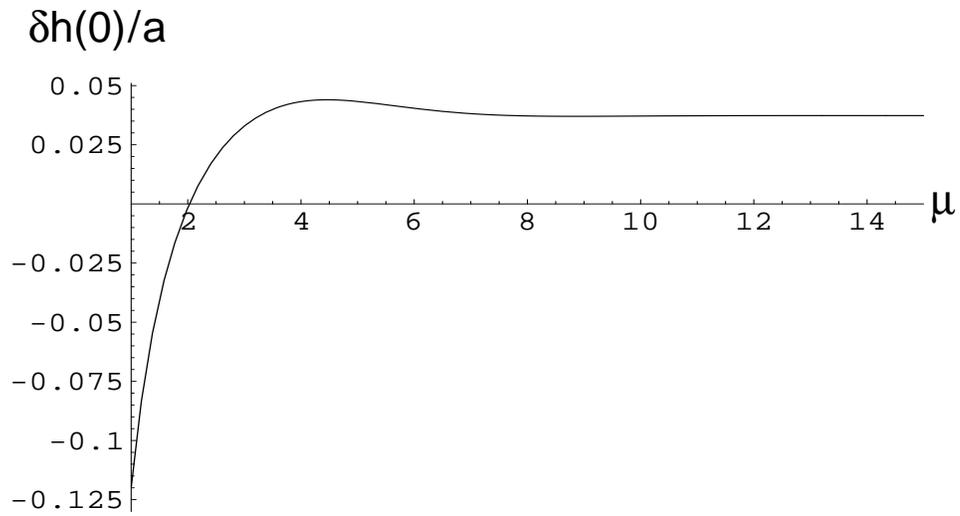}}
\end{center}
\caption{The height deviation of the membrane in the middle of a trench
(strong adhesion), $\delta h(0) = h(0)-h_0$, as a function of $\mu$ for
fixed $\lambda=0.12$ and $d=2a\simeq 99$\AA.}
\label{fig11}
\end{figure}

Finally, in Fig.\ \ref{fig11} we plot $\delta h(0)$ against the
periodicity parameter, $\mu$. It is, perhaps, at first sight surprising
to see that for large $\mu$, $\delta h(0)$ does not vanish. However, this
is the wrong function with which to take this limit, as one always has at
least one trench in the system and is always looking at the depth in this
trench. If instead $\delta h(\mu d/2)$ is considered, this tends quite
rapidly to $\lambda d/2$. For example, with the parameters given in Table
\ref{tab} and Fig.\ \ref{fig11}, $\delta h(d) \simeq 0.120a$ when
$\mu=25$. Hence, the membrane returns to its height above a planar
substrate. However, one can see from Fig.\ \ref{fig11} that for $\mu
\simeq 8$, the trenches no longer have a significant effect
 and $\delta h(0)$ plateaus; the membrane effectively experiences a
potential generated by a flat surface.

\subsection{Self-affine surfaces}
\label{selfaffine}

In this penultimate section, we describe surfaces whose roughness is
random in character. This is normally used to model the ``natural''
roughness of a substrate which arises due to chemical impurities or
crystal defects. We adopt the self-affine approximation
\cite{vol14} and assume that the substrate appears relatively smooth on
large lengthscales. Typical fluctuations of length $L$ grow only as $L
^\beta$ where $\beta$ is the roughness exponent $0 \le \beta \le 1$,
which is directly analogous to the thermal roughness of an interface
\cite{lipfish}. However, such models become divergent as $L \goto \infty$
and we prefer to adopt a reduced self-affine scaling law \cite{stan}
where a long lengthscale cut-off has been explicitly included
\equ
\langle |z_\rw(\y' + \y)-z_\rw(\y')|^2 \rangle = 2 \gamma^2 \left(
1-\e^{-\left( \frac{\rho}{\zeta} \right)^{2\beta}} \right) 
\uqe
or equivalently
\equn{sa}
\langle z_\rw({\bf 0}) z_\rw({\y}) \rangle = \gamma^2 \e^{-\left(
\frac{\rho}{\zeta} \right)^{2\beta}}
\nuqe
Here $\gamma$ is the saturated root mean square roughness at long
lengthscales, and $\zeta$ is the crossover length from self-affine
 to saturated behavior. The wetting properties of such surfaces
($\ka=0$) have already been quite extensively studied \cite{dis,david2}.

To be able to calculate the membrane adhesion energy analytically, we
specialize to two cases: 

\subsubsection{Roughness $\beta = 1$}

Equation (\ref{sa}) can be Fourier transformed to give
\equn{fb1}
\langle \tilde{z}_\rw({\bf p}) \tilde{z}_\rw(\q) \rangle = 4 \pi^3 \gamma^2
\zeta^2 \e^{-\frac{\zeta^2 q^2}{4}} \delta({\bf p} + \q) 
\nuqe
which can then be used in (\ref{Udery}) to calculate $\Delta U$. We find
that
\equn{sa1}
\frac{\Delta U}{\Si} = - \frac{1}{8} \left( \frac{\gamma {\cal X}^2}{\xi}
\right)^2 \Biggl\{ 4 + \left( \frac{\zeta {\cal X}}{\xi} \right)^4 \left[
\frac{I(\Lambda_{+}) - I(\Lambda_{-})}{\Lambda_{+}^2-\Lambda_{-}^2} \right]
\Biggr\}
\nuqe
where
\eqa
I(u) &=& \e^\frac{u^2}{4} \Gamma \left( 0,\frac{u^2}{4} \right) \\
\Lambda_\pm &=& \zeta \eta_\pm 
\aqe
and 
\equ
\Gamma(0,u) = \int_u^\infty dz \frac{\e^{-z}}{z}
\uqe 
is an incomplete Gamma function.

\subsubsection{Roughness $\beta = \half$}

In this case we have
\equ
\langle \tilde{z}_\rw({\bf p}) \tilde{z}_\rw(\q) \rangle = \frac{2 \pi^3
\gamma^2 \zeta^2}{(1+\zeta^2 q^2)^\frac{3}{2}} \delta({\bf p} + \q)
\uqe
and
\equn{sa2}
\frac{\Delta U}{\Si} = -\frac{1}{8} \left( \frac{\gamma {\cal X}^2}{\xi}
\right)^2 \Biggl\{ 1 + \left(\frac{\zeta {\cal X}}{\xi} \right)^4 \left[
\frac{I(\Theta_{+})-I(\Theta_{-})}{\Theta_{+}^2-\Theta_{-}^2} \right]
\Biggr\}
\nuqe
with
\equ
I(u) = \frac{1}{u^2} + \frac{1}{2u^3} \log 
\left( \frac{1-u}{1+u} \right) 
\uqe
and 
\equ
\Theta_\pm^2 = 1 -(\zeta \eta_\pm)^2 
\uqe

\begin{figure}
\begin{center}
\scalebox{0.6}{\includegraphics{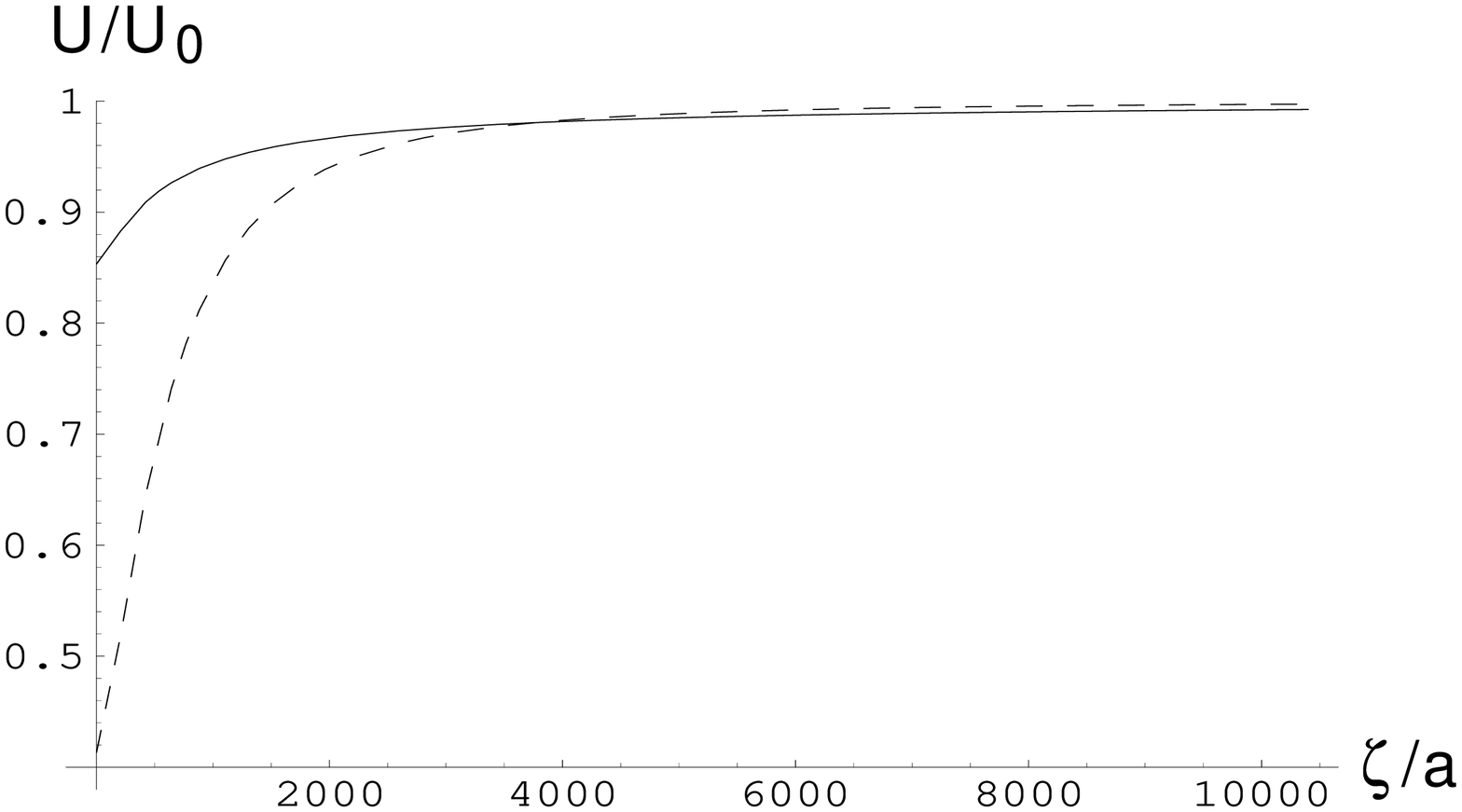}}
\end{center}
\caption{The relative adhesion energy for a weakly adhering membrane above
two different self-affine surfaces plotted against $\zeta$, the crossover
length; the full curve is for $\beta=\half$ while the dashed is
$\beta=1$. Here, $\gamma = 4a$, that is $\gamma \simeq 114 \; {\mbox \AA}$
and all other parameter values are given in Table \ref{tab} (weak
adhesion).}
\label{fig12}
\end{figure}

For $\gamma=4a$, equations (\ref{sa1}) and (\ref{sa2}) are illustrated in
Fig.\ \ref{fig12}. As the crossover length, $\zeta$, grows, the disorder
becomes more and more correlated and the substrate becomes smoother on
larger lengthscales. One can see that in both cases, $U$ is tending
towards the planar value. For smaller $\beta$, this plateau is reached
more quickly in accordance with (\ref{sa}). As $\zeta$ decreases and the
substrate becomes rougher, the adhesion energy deviates further and
further away from the planar value. 

\subsection{Tensionless membranes}

So far we have only considered membranes which have a finite tension, $\Si$,
 and now, for the sake of completeness, we will briefly illustrate how our
results change in the limit of $\Si \goto 0$. For vanishing membrane
tension, the definitions (\ref{xidefn}, \ref{lams}) and (\ref{K}) imply
that
\equn{zeroetas}
\eta_\pm = \pm \frac{1}{\sqrt{2}} (1\pm i) \xi_\ka^{-1} \hspace*{2cm}
\mbox{for $\Si=0$}
\nuqe
Equation (\ref{gst}) then indicates that the profile will have an
oscillatory though exponentially damped form. 

To see how the adhesion energy changes, we can look at, for example, the
adhesion energy of Sec.\ \ref{trsSec}, equation (\ref{treU}) which is
valid for a surface periodically patterned with trenches. Multiplying
(\ref{treU}) by $\Si$, one is naturally led to consider
\equ
\frac{\Si \chi^4}{\xi^2} \goto \ka \xi_\ka^{-4}
\uqe
in the limit of zero tension. A little thought then gives
\equ
\Delta U = \frac{\lambda^2}{\mu} \left( \frac{d}{\xi_\ka} \right)^2 \ka
\xi_\ka^{-2} \Biggl[ I(\Lambda_{+},\Lambda_{-}) +
I(\Lambda_{-},\Lambda_{+}) \Biggr]
\uqe
with
\equ
I(u,v) = \frac{(\e^u-1)(\e^{(\mu-1)u}-1)}{\e^{\mu u}-1} \cdot
\frac{u v^2}{(u^2-v^2)^2} 
\uqe
and $\Lambda_\pm = d \eta_\pm$, with the $\eta$'s obeying
(\ref{zeroetas}). While it is not as obvious as before, $\Delta U$ is
again negative. 


\section{Discussion}
\setcounter{equation}{0}

In this paper, we have looked at the effect of geometric surface
structure on the adhesion properties of a membrane and a solid substrate.
The membrane interacts with the substrate via a Van der Waals potential
and experiences either a Helfrich-like entropic repulsion, if it is weakly
adhering, or hydration forces, if it is strongly adhering or supported. We
find that the most analytically successful approach is that of Deryagin
in which the entire interaction potential is assumed to be a local
function of height. By making a further harmonic approximation quite a
large number of surface geometries can be solved for. In general, the
membrane tries to follow the surface contour of the substrate but for
short wavelength undulations is damped due to its tension and
rigidity. Two lengthscales, $\xi_\Si$ and $\xi_\ka$, analogous to the
``healing'' length \cite{david} for interfaces, emerge and control this
damping.

Within the Deryagin description, we find that the membrane can overshoot
(similar to the behavior in a laser trap \cite{barziv}) near a
disturbance in the planar structure of the surface. Overshooting (i.e\ having
a height greater than that it would have had for a completely flat
surface)  is seen to be a consequence of the eigenvalues of the 4$^{th}$
order operator, making up the Euler-Lagrange equation, going complex.
This occurs for $2\xi > \xi_\Si$ and consequently is directly due to the
membrane having a finite (non-zero) rigidity.

In general (but still at the Deryagin level), increasing the roughness of
the surface reduces the membrane adhesion's energy. Having a non-planar
substrate immediately leads to a competition between the bending and
potential energy contributions to the free energy; the bending energy
tries to flatten the membrane and so move it away from the surface to
heights where the effect of the geometrical heterogeneity is washed out.
On the other hand, the potential energy, which, due to the larger surface
area of the substrate, is more attractive than in the planar case, acts
to bring the membrane in closer. As the elastic or bending energy of a
membrane contains a rigidity term it always dominates over the potential
energy (for a membrane exhibiting a sharp kink configuration this
rigidity contribution even diverges -- see Appendix) and thus the
adhesion energy is reduced. For a flat substrate, the bending energy
contribution is zero and this rivalry does not exist. 

This inverse dependence of $U$ on the roughness of the substrate is
perhaps best illustrated in Fig.\ \ref{fig10}, where by changing $\mu$ we
move through different surface configurations whose roughness function
(the variance of $z_\rw$) passes through a minimum. The adhesion energy
can be clearly seen to mimic such behavior. In Fig.\ \ref{fig4}, the
importance of the bending energy is again shown; with $U$ decreasing as
the rigidity, $\ka$, increases.

Figures \ref{fig6} and \ref{fig11} describe the extent to which the
membrane penetrates into an indentation in the substrate surface.
Generally, the membrane extends more extensively into a wider trench; the
bending energy of the resulting configuration is less --- the membrane is
flatter, and the potential energy gain is also greater as the membrane is
more able to straighten out at the trench base. 

As mentioned in the introduction, supported membranes are invaluable
for the construction of biosensors. The proteins that endow the membrane
with this properties can, unfortunately, disturb it from a favorable
planar configuration. One way to surmount  this problem is to indent the
substrate with pockets which then act as ``docking'' pods for the
proteins. However, this technique can backfire if the membrane itself
changes its configuration significantly in response to the now non-planar
surface. Our theoretical work predicts that the narrower the trench or
pocket, Fig.\ \ref{fig6}, and the more widely spaced apart they are,
Fig.\ \ref{fig11}, the less likely such an event occurs. We also find
that, within the Deryagin framework, increasing the depth of the
trenches encourages the membrane to penetrate further into them.

Finally, we looked at randomly rough or self-affine surfaces where the
general principles outlined above still hold. Rougher surfaces (those
with a larger $\beta$ exponent) have smaller adhesion energies, see Fig.\
\ref{fig12}.

In an accompanying paper \cite{II}, we will explore non-local methods,
which enable progress beyond the Deryagin approximation, and also will
perform some exact numerical calculations. We intend to extend the method
to include chemical heterogeneity. This then opens up the possibility of
modeling chemically {\it and} geometrically structured substrates which
are becoming crucial to recent pioneering biotechnological research.

\section*{Acknowledgments}

We are particularly grateful to J.\ R\"adler and E.\ Sackmann for
introducing us to the problem of membrane adhesion on rough surfaces, for
numerous discussions and suggestions, and for sharing with us their
experimental results.  We greatly benefited from conversations and
correspondence with M.\ Kozlov, A.\ Marmur, P.\ Lenz, R.\ R.\ Netz, M.\ O.\
Robbins, U.\ Seifert and P.\ B.\ Sunil Kumar.  PSS would like to thank
the British Council, the Israeli Ministry of Science and the Sackler
Institute for Theoretical Physics (Tel Aviv University)  for providing
financial support during a visit to Israel, where the majority of this
work was carried out.  Partial support from the Israel Science Foundation
founded by the Israel Academy of Sciences and Humanities --- centers of
Excellence Program and the U.S.-Israel Binational Foundation (B.S.F.) 
under grant No. 94-00291 is gratefully acknowledged.

\section*{Appendix: Kink configurations in membranes and interfaces}
\setcounter{equation}{0}

In this appendix, we consider the elastic energy cost of a membrane
adopting a configuration containing sharp kinks.  Let us, for the purpose
of illustration, consider a one-dimensional membrane embedded in a
two-dimensional space having the profile
\equ
h(x) = c \tanh \left( \frac{x}{\epsilon L} \right)
\uqe
with the two lengths $L$ and $c$ obeying $L \gg c$. In the limit of
$\epsilon \goto 0$, $h(x)$ develops two sharp kinks
\equ
\lim_{\epsilon \goto 0} h(x) = c [2\theta(x)-1]     
\uqe
and becomes a step function. We can examine the effect this has on the
asymptotic behaviors of the two adhesion energy densities (per unit
area) 
\equn{cont1}
u_\Si = \frac{\Si}{2L} \int_{-L}^L dx \, \sqrt{1+h^{'2}}
\nuqe
and
\equn{cont2}
u_\ka = \frac{\ka}{2L} \int_{-L}^L dx \,
\frac{h^{''2}}{(1+h^{'2})^\frac{5}{2}}
\nuqe
Equations (\ref{cont1}) and (\ref{cont2}) provide the contributions to
the free energy from the tension (interface-like) and the rigidity
(membrane-like) terms in (\ref{f}), respectively. 

The interfacial elastic energy cost is simply proportional to the length
of the membrane in the step configuration
\eqa
u_\Si & \simeq & \frac{\sigma}{2L} \cdot 2(L+c) \nonu \\
& \sim & \sigma
\aqe
where we have taken the limit of $\epsilon \goto 0$. This is clearly finite.

To find the asymptotic limit of the rigidity contribution, we notice that
\equ
u_\ka = \frac{2 c^2 \epsilon \ka}{L^5} \int_{-L}^L dx \, \frac{\tanh^2
\left( \frac{x}{\epsilon L} \right) \cosh^{-4}\left( \frac{x}{\epsilon L}
\right)}{\left[ \epsilon^2 + \left( \frac{c}{L} \right)^2 \cosh^{-4}
\left( \frac{x}{\epsilon L} \right) \right]^\frac{5}{2}}
\uqe
is an even function and so, writing $t=\tanh \left( \frac{x}{\epsilon L}
\right)$, we have
\equ
u_\ka = \frac{4 c^2 \ka}{\epsilon^3 L^4} \int_0^{\tanh(1/\epsilon)} dt \,
\frac{t^2(1-t^2)}{\left[ 1+ \left( \frac{c}{\epsilon L} \right)^2
(1-t^2)^2 \right]^\frac{5}{2}}
\uqe
which can be integrated by parts
\equn{Aine}
u_\ka \sim \frac{2\ka}{3L^2\epsilon} \left\{ 1 -
\int_0^{\tanh(1/\epsilon)} \frac{dt}{\left[ 1+ \left( \frac{c}{L}
\left(\frac{1-t^2}{\epsilon}\right) \right)^2 \right]^\frac{3}{2}}
\right\}
\nuqe
As 
\equ
\lim_{\epsilon \goto 0} \left\{ \frac{1-\tanh^2(1/\epsilon)}{\epsilon}
\right\} = 0
\uqe
the integral in (\ref{Aine}) is well-defined and finite.
 
Therefore, for small $\epsilon$, we have 
\equ
\begin{array}{lcr}
u_\Si \sim \Si & \hspace*{5mm} {\rm and} \hspace*{5mm} & u_\ka \sim
\frac{\ka}{L^2} \epsilon^{-1}
\end{array}
\uqe
and so the tension or interface-like contribution is constant while that
from the rigidity (membrane-like)  diverges -- an interface ($\ka=0$ and
$\Si \ne 0$) can have a configuration with a sharp kink while a membrane
cannot. 


\pagebreak

\newpage
\begin{table}[ht]
\renewcommand{\arraystretch}{1.5}
\renewcommand{\tabcolsep}{1.4mm}
\begin{center}
\footnotesize{
\begin{tabular}{llll}
\hline
\hline
$\ka = 35T$ & $\Si = 1.7 \times 10^{-5} \; {\rm Jm}^{-2}$ 
& $\delta = 38\; {\mbox \AA}$ & $p = 0$ \\ 
$\Omega \simeq 8.07\times 10^7 \; {\rm m}^{-1}$ & 
$b \simeq 5.47\times 10^4 $ & $\alpha^{-1} =
2.2 \; {\mbox \AA} $& $T = 4.1 \times 10^{-21}\; {\rm J}$\\
\hline
& WEAK ADHESION & &\\
\hline 
$A = 8.67 \times 10^{-22} \; {\rm J}$ & $a \simeq 28.5 \; {\mbox \AA}$ &
$h_0 \simeq 11.85 a\simeq 338\; {\mbox \AA}$ 
& $U_0 \simeq 1.1 \times 10^{-4} \Si$ \\ 
$\xi \simeq 32.25a$ & $v \simeq 8.04 \times 10^{-6} a^{-2} \Si $ & $ \xi_\Si
\simeq 352.62 a $ & $\xi_\ka \simeq 106.64 a $ \\ 
$\chi \simeq 0.302$
& $\eta_+ \simeq 0.0309a^{-1}$ & $\eta_- \simeq 0.0028 a^{-1}$ & $L
\simeq 259.1 a$ \\
\hline
&STRONG ADHESION &(supported membrane)  \\ 
\hline
$A = 2.6 \times 10^{-21} \; {\rm J}$ & $a \simeq 49.3 \; {\mbox \AA}$ &
$h_0 \simeq 0.61 a\simeq 30\; {\mbox \AA}$ 
& $U_0 \simeq 0.298 \Si$ \\ 
$\xi \simeq 18.62a$ &
$v \simeq 22.85 a^{-2} \Si $ & $ \xi_\Si \simeq 0.21a $ & $\xi_\ka
\simeq 1.97 a $ \\ 
$\chi \simeq 9.44$ & $\eta_+ \simeq (0.359 + 0.357
i)a^{-1}$ & $\eta_- \simeq (0.359-0.357i) a^{-1}$ & $L \simeq 22.3 a$
\\
\hline
\hline
\end{tabular}
}
\end{center}
\caption{The various parameters, chosen and calculated, for the planar
system and for both weak and strong adhesion. For definitions see text.}
\label{tab}
\end{table}

\newpage

\newpage

\newpage

\newpage

\newpage

\newpage

\newpage

\newpage

\newpage 

\newpage

\newpage

\newpage


\begin{thebibliography}{99}

\bibitem{sackmann} See, for example, {\it Structure and Dynamics of
Membranes}; Lipowsky R.\ and Sackmann E., Eds.; Elsevier: Amsterdam, 1995. 

\bibitem{alberts} Alberts B., Bray D., Johnson A., Lewis J., Raff M.,
Roberts K.\ and Walter P. {\it Essential Cell Biology}; Garland
Publishing: New York, 1998.

\bibitem{lasic} Lasic D.\ D. In {\it Structure and Dynamics of
Membranes}; Lipowsky R.\ and Sackmann E., Eds.; Elsevier: Amsterdam, 1995;
p 491--519. 

\bibitem{seiflip} Seifert U.\ and Lipowsky R. Phys.\ Rev.\ A {\bf 1990},
{\it 42}, 4768. Kraus M., Seifert U.\ and Lipowsky R. Europhys.\ Lett.\
{\bf 1995}, {\it 32}, 431.

\bibitem{seif} Seifert U. Phys.\ Rev.\ Lett.\ {\bf 1995}, {\it 74}, 5060. 

\bibitem{rad} R\"{a}dler J.\ O., Feder T.\ J., Strey H.\ H.\ and Sackmann
E. Phys.\ Rev.\ E {\bf 1994}, {\it 51}, 4526. 

\bibitem{sack} Sackmann E. Science {\bf 1996}, {\it 271}, 43. R\"{a}dler
J.\ and Sackmann E. Curr.\ Opin.\ Solid State Material Sci.\ {\bf 1997},
{\it 2}, 330.

\bibitem{sala} Salafsky J., Groves J.\ T.\ and Boxer S.\ G.  Biochemistry
{\bf 1996}, {\it 35}, 14773.

\bibitem{gerdes} Gerdes S.\ and Str\"{o}m G. Colloids Surfaces A {\bf
1996}, {\it 116}, 135.

\bibitem{moller} Spatz J.\ P., Herzog T., M\"{o}ssmer S., Ziemann P.\ and
M\"{o}ller M. Adv.\ Mater.\ {\bf 1999}, {\it 11}, 149.

\bibitem{marmur} Wolansky G.\ and Marmur A. Langmuir {\bf 1998}, {\it
14}, 5292.

\bibitem{swalip} Swain P.\ S.\ and Lipowsky R. Langmuir {\bf 1998}, {\it
14}, 6772. 

\bibitem{dis} Kardar M.\ and Indekeu J.\ O. Europhys.\ Lett.\ {\bf 1990},
{\it 12}, 161. Li H.\ and Kardar M. Phys.\ Rev.\ B {\bf 1990}, {\it 42},
6546. Sartoni G., Stella A.\ L., Giugliarelli G.\ and Dorsogna M.\ R.
Europhys.\ Lett.\ {\bf 1997}, {\it 39}, 633. 

\bibitem{sp} Parry A.\ O., Swain P.\ S.\ and Fox J.\ A. J.\ Phys.: 
Condens.\ Matter {\bf 1996}, {\it 8}, L659. Swain P.\ S.\ and Parry A.\
O.  Eur.\ Phys.\ J.\ B {\bf 1998}, {\it 4}, 459. 

\bibitem{peter} Gau H., Herminghaus S., Lenz P.\ and Lipowsky R.  Science
{\bf 1999}, {\it 283}, 46. 

\bibitem{david} Andelman D., Joanny J.\ F.\ and Robbins M.\ O. 
Europhys.\ Lett.\ {\bf 1988}, {\it 7}, 731. Robbins M.\ O., Andelman D.\
and Joanny J.\ F. Phys.\ Rev.\ A {\bf 1991}, {\it 43}, 4344.

\bibitem{dr} Netz R.\ R.\ and Andelman D. Phys.\ Rev.\ E {\bf 1997}, {\it
55}, 687. 

\bibitem{david2} Harden J.\ L.\ and Andelman D. Langmuir {\bf 1992}, {\it
8}, 2547.

\bibitem{II} Swain P.\ S.\ and Andelman D., to be published. 

\bibitem{canham} Canham P.\ B.  J.\ Theoret.\ Biol.\ {\bf 1970}, {\it
26}, 61. Helfrich W. Z.\ Naturforsch.\ C {\bf 1973}, {\it 28}, 693.

\bibitem{lip} Lipowsky R. In {\it Structure and Dynamics of Membranes};
Lipowsky R.\ and Sackmann E., Eds.; Elsevier: Amsterdam, 1995; p 521--602.

\bibitem{isr} Israelachvili J.\ N. {\it Intermolecular and Surface
Forces}; Academic Press: London, 1992. 

\bibitem{lipfish} Lipowsky R.\ and Fisher M.\ E. Phys.\ Rev.\ B {\bf
1987}, {\it 36}, 2126.

\bibitem{helf2} Helfrich W. Z.\ Naturforsch.\ {\bf 1978}, {\it 33}, 305. 

\bibitem{helfserv} Helfrich W.\ and Servuss R.\ M. Il Nuovo Cimento D
{\bf 1984}, {\it 3}, 137.

\bibitem{eva} Evans E., Langmuir {\bf 1991}, {\it 7}, 1900. 

\bibitem{rolandlip} Netz R.\ R.\ and Lipowsky R. Europhys.\ Lett.\ {\bf
1995}, {\it 29}, 345.
 
\bibitem{bruinsma} Bruinsma R., Goulian M.\ and Pincus P. Biophys.\ J.\
{\bf 1994}, {\it 67}, 746.

\bibitem{barziv} Bar-Ziv R., Menes R., Moses E.\ and Safran S.\ A. 
Phys.\ Rev.\ Lett.\ {\bf 1995}, {\it 75}, 3356. 

\bibitem{space} Menes R.\ and Safran S.\ A. Phys.\ Rev.\ E {\bf 1997},
{\it 56}, 1891. 

\bibitem{roland} Netz R.\ R. J.\ Phys.\ (France) I {\bf 1997}, {\it 7},
833.

\bibitem{lipleib} Lipowsky R.\ and Leibler S. Phys.\ Rev.\ Lett.\ {\bf
1986}, {\it 56}, 2541.

\bibitem{pars} Gawrisch K., Ruston D., Zimmerberg J., Parsegian V.\ A.,
Rand R.\ P.\ and Fuller N. Biophys.\ J.\ {\bf 1992}, {\it 61}, 1213.

\bibitem{pgg} de Gennes P.\ G. Rev.\ Mod.\ Phys.\ {\bf 1985}, {\it 57},
827. 

\bibitem{shirl} Johnson S.\ J., Bayerl T.\ M., McDermott D.\ C., Adam G.\
W., Rennie A.\ R., Thomas R.\ K.\ and Sackmann E. Biophys.\ J.\ {\bf
1991}, {\it 59}, 289.

\bibitem{dery1} Deryagin B.\ V., Kolloidn.\ Zh.\ {\bf 1955}, {\it 17},
827. Deryagin B.\ V., Churaev N.\ V.\ and Muller V.\ M. {\it Surface
Forces}; Consultants Bureau: New York, 1987.

\bibitem{warning} Assuming (\ref{perturb}) is still valid. 

\bibitem{over} Similarly to the exponentials in (\ref{gst}), $K_0(\eta)$
and $I_0(\eta)$ are single valued functions for real $\eta>0$ and so an
overshoot can only occur for complex $\eta$. 

\bibitem{vol14} See, for example, the review article by Forgacs G.,
Lipowsky R.\ and Nieuwenhuizen Th.\ M. In {\it Phase Transitions and
Critical Phenomena}; Domb C.\ and Lebowitz J.\ L., Eds.; Academic Press:
London, 1991; Vol.\ 14. 

\bibitem{stan} Sinha S.\ K., Sirota E.\ B., Garoff S.\ and Stanley H.\ B.
Phys.\ Rev.\ B {\bf 1988}, {\it 38}, 2297. 

\end{thebibliography}
\end{document}